\documentclass[aps,pre,reprint, superscriptaddress]{revtex4-2}
\usepackage{blindtext}
\usepackage[colorlinks=true,allcolors=blue]{hyperref}

\usepackage{blindtext}
\usepackage{bm}

\def \beq{\begin{eqnarray}}
\def \eeq{\end{eqnarray}}

\def \br{\bm{r}}

\usepackage{amsmath}
\usepackage{amssymb}
\usepackage{graphicx}
\usepackage{graphicx,xcolor}

\usepackage{amsmath}

\definecolor{nblue}{RGB}{28,130,185}
\definecolor{cgreen}{RGB}{76,153,0}
\definecolor{myorange}{RGB}{245,156,74}

\usepackage{hyperref}
\hypersetup{
  colorlinks=true,
  citecolor=cyan,
  urlcolor=cgreen
}

\def \beq{\begin{eqnarray}}
\def \eeq{\end{eqnarray}}

\begin{document}

\title{Compositional disorder in a multicomponent nonreciprocal mixture: stability and patterns}
\author{Laya Parkavousi}
\email{laya.parkavousi@ds.mpg.de}
\author{Suropriya Saha}
\email{suropriya.saha@ds.mpg.de}
\affiliation{Max Planck Institute for Dynamics and Self-Organization (MPIDS), D-37077 G\"ottingen, Germany}

\date{\today}

\begin{abstract}
In passive phase-separating mixtures, the average density of each component can be tuned to control the composition of the coexisting bulk phases. This concept extends to active systems. For example, in a nonreciprocal mixture of two species, changing the average density of either species alters the qualitative nature of the travelling patterns that emerge in the steady state. In this paper, we build on the existing framework of the multi-species nonreciprocal Cahn–Hilliard (NRCH) model to examine the influence of compositional disorder in a multicomponent active system. Specifically, we allow each scalar field in the mixture to have a distinct average density and analyze the implications of this generalized setting. Focusing on ensembles of systems in which the inter-species interaction coefficients and the average densities of each species are both sampled from probability distributions, we show that nonreciprocity stabilizes the uniformly mixed state, even in the presence of compositional disorder, when compared to the corresponding reciprocal system. Using random matrix theory, a general condition for the onset of the spinodal instability is derived and verified through simulations. Finally, the connection between the statistics of the most unstable eigenvalue and the emergent nonlinear dynamics is illustrated.

\end{abstract}

\maketitle
\section{Introduction} 
\noindent
An important question that has driven intense research is -- how does a cell achieve self-organization and regulation? A widely accepted route, applicable in many scenarios, is that the constituents of the cell cytoplasm organize into compartments enriched in a specific group of chemicals simply through thermodynamic phase separation~\cite{CliffBrangwynne_2009, Alberti2021}. In the past decade, liquid-liquid phase separation has been implicated in the formation of membrane-less organelles~\cite{HymanReview, shin2017liquid, cohan2020making, brangwynne2015polymer}, regulation of stable structures such as the nucleolus~\cite{Iarovaia2019, brangwynne2011active, lafontaine2021nucleolus}, and in strategies to adapt to adverse external stimuli in stress granules~\cite{Protter2016, Franzmann2018}.  It is striking that this intuitively appealing idea that has recently received widespread acceptance, surfaced for the first time at the end of the nineteenth century to explain the shape of sea-urchin protoplasm~\cite{EdmundWilson_1899,Wang2021} and again in the context of proto-cells which might have paved the way for the origin of life~\cite{oparin2003origin}.  The identification of phase separation as the dominant mechanism opens up new possibilities for regulating condensate formation through pathways accessible in passive systems. These include controlling pH~\cite{Adame-Arana2020}, modulating chemical activity~\cite{Patel_ATP_LLPS}, applying a chemical gradient~\cite{Weber_2017}, coupling to an active fluid~\cite{Tayar2023}, and, most relevant to this work, altering the relative mean density of the constituents~\cite{PeterSollich_2002}. The exploration of the last pathway, namely, controlling the total availability of a particular chemical species in an active phase-separating mixture, is the main topic of this paper.
 \begin{figure*}
	\centering
	\includegraphics[width= 0.95\linewidth]{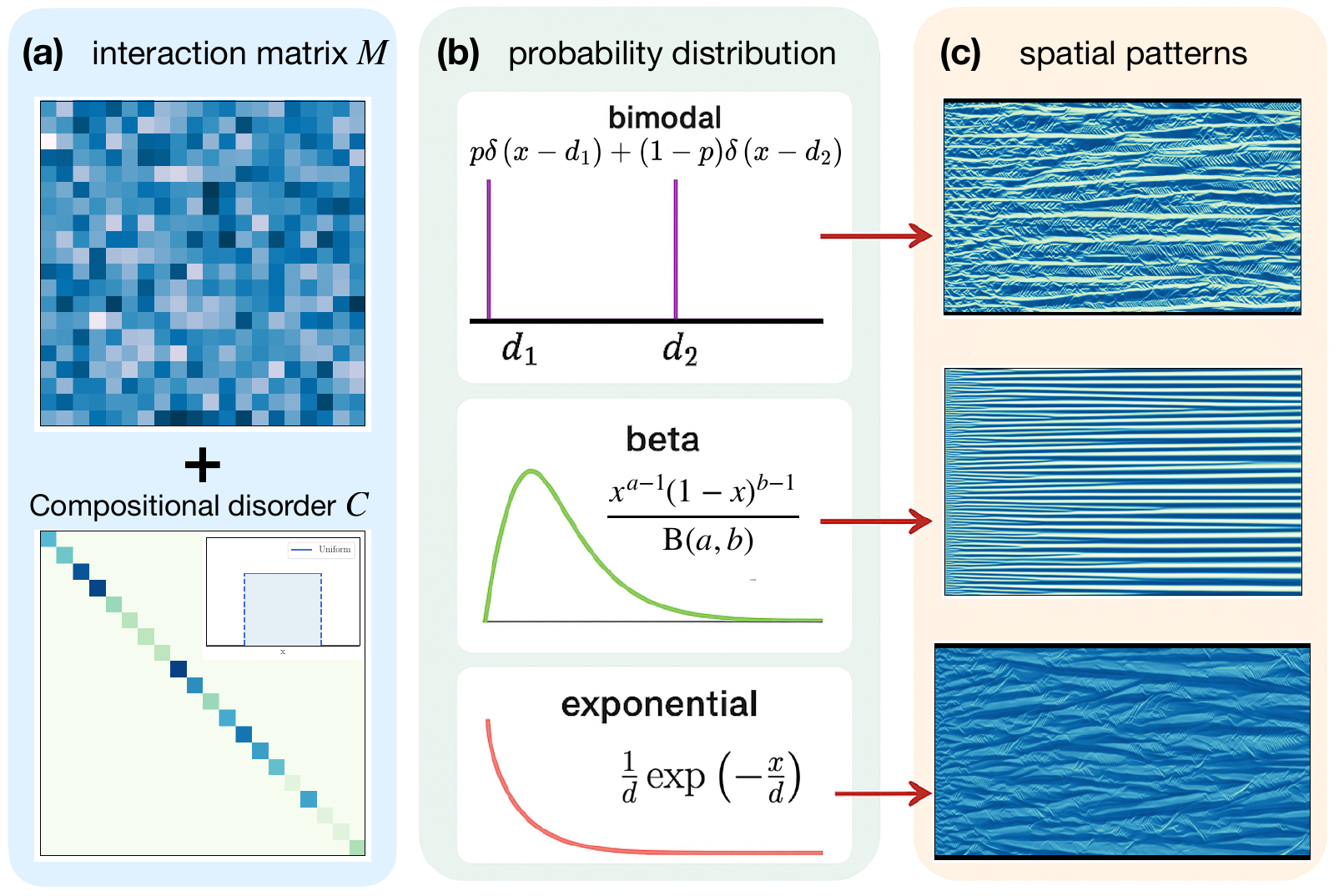}
	\caption{\textbf{Role of compositional disorder in multi-species NRCH}: (a) The average composition of each species is fixed to a value that is drawn from a distribution. The dynamical matrix determining the linear stability of the homogeneous mixed state is the sum of a random Gaussian matrix (top) and a diagonal matrix whose entries are determined by the mean density of each species (bottom). In (a) the diagonal entries are drawn from a uniform distribution and each color shows a different random value. (b) Three other distributions are chosen to illustrate our main results - the associated probability distribution functions and their name is mentioned. (c) To illustrate the effect of compositional disorder on the dynamics, we run simulations in one dimensional space keeping $\mathbb{M}$ the same in all three cases. The top and the bottom evolve to a state of spatiotemporal chaos in the steady state, while the middle evolves to a phase separated state. }
	\label{fig:Schematic}
\end{figure*}

A description of processes within the cell cytoplasm has to incorporate two features: multiple participants~\cite{zimmermann1997} and activity in a suitable form. A recent protein database~\cite{You2020} lists more than two thousand relevant types of proteins in membrane-less organelles that are collected from hundreds of papers. The heterogeneity at the cellular level is reflected in numerous studies on passive phase separation~\cite{Sollich2001, Sear_2003, jacobs2023theory, jacobs2017phase,Shrinivas2021}. However, the perspective that active processes must be included in a suitable, system-specific form in a theoretical description of bio-condensates is relatively recent~\cite{WeisHondele2020,aierken2026roadmapcondensatescellbiology}, such that the properties of active mixtures constitute a new area of research. In the field of active matter, active phase separation in a single component has been explored in several minimal models, such as motility-induced phase separation~\cite{cates2015motility, Wittkowski2014}, chemically active mixtures~\cite{Rabea-Natphy-2017, Brauns_Frey_PhysRevX.10.041036}, and binary mixtures of nonreciprocally interacting components~\cite{saha2020, Aparna-non-reciprocity-PNAS-20}.

With this background, we assume a reductionist approach where we model the cell as a multicomponent mixture with nonreciprocal inter-species interactions~\cite{soto2014, soto2015, saha2019, OuazanReboul2023, parkavousi2024enhanced, dinelli2023non, loos2020irreversibility}. Nonreciprocity appears naturally in mixtures of active particles: whether the constituents have an orientation~\cite{fruchart2021} or not~\cite{saha2020,Aparna-non-reciprocity-PNAS-20}, in quorum sensing mixtures~\cite{BenoitYu2023,dinelli2023non}, or mixtures of Janus colloids~\cite{tucci2024nonreciprocal}. Nonreciprocity is a key ingredient that can be used to mimic behaviors typically associated with living, intelligent systems, such as predator–prey dynamics. Recent experimental findings have established the validity of some key theoretical predictions in polar mixtures~\cite{chen2024emergent}, active solids~\cite{guillet2025melting,tan2022odd,chao2026selective}, and a vibrated granular mixture of rod and spheres~\cite{SriramExptNRCH_PhysRevLett.133.208301}.  

The hallmark of nonreciprocal number conserving systems is that it combines features of pattern-forming systems~\cite{saha2022, Pisegna2024} and phase separating systems~\cite{saha2024phasecoexistencenonreciprocalcahnhilliard,ThielePS}. The latter aspect implies that the average density is an important physical quantity that can be tuned to alter dynamics, as shown explicitly for a binary nonreciprocal mixture where the system undergoes a transition from stable mixtures to traveling lattices via a conserved form of the Hopf bifurcation~\cite{saha2020}. The role of composition is, however, largely unexplored in these systems, with a few exceptions in the binary case~\cite{saha2020}, and no studies in the multicomponent case.

In this work, we address a natural question: how does variability in the mean density of each species, hereafter referred to as compositional disorder, influence the stability and pattern formation in the system? Recent work~\cite{parkavousi2024enhanced} has shown that nonreciprocity, in the form of asymmetric correlations of inter-species interaction coefficients, stabilizes the homogeneous mixed state compared with a system that has only passive interactions. If the mean density of each species is identical, its effect can be subsumed into the definition of a temperature-like control parameter that can be tuned to drive pattern formation in the active system. The simplifying assumption, however, is invalid for biological condensates, as established in recent papers where it has been possible to measure the relative abundance of each component forming a condensate and show that it varies widely~\cite{McCall2025}, implying that a realistic description of the underlying physics should account for this variability. In biological condensates the compositions are typically highly variable, for example in cancer cells \cite{Klein2020} and in experiments using DNA chains to explore this diversity in a controlled manner~\cite{chaderjian2025diversedistinctdenselypacked}.

We use random matrix theory~\cite{schehr2017, mehta2004, bordenave2012around, bai2010spectral} to study the linear stability of the system, as pioneered in the seminal works of Wigner~\cite{wigner2} and May~\cite{MAY1972} in quantum mechanics and ecosystems, respectively. This approach has been successfully applied to a wide range of physical systems, including the stability of ecosystems with random interactions~\cite{mougi2012diversity, altieri2022effects, biroli2018marginally, baron2022eigenvalues, baron2020dispersal}, and disordered systems such as spin glasses~\cite{mezard1987spin, crisanti1988dynamics, crisanti1987dynamics}. We adopt a statistical perspective, replacing the complexity of individual interaction networks with probabilistically sampled ensembles of systems~\cite{Sear_2003, Liquid_Hopfield, dinelli2025randommotilityregulationdrives}. Within this framework, the mean composition is treated as a stochastic variable drawn from a distribution, as recently investigated in passive systems~\cite{thewes2023composition}. The stability analysis of uniform mixtures of the multi-species NRCH model reduces to studying the eigenvalue spectrum of the asymmetric matrices with diagonal disorder (diagonal entries being pulled from a different, non-Gaussian distribution) as represented in Fig.~\ref{fig:Schematic}(a).  The approach facilitates the statistical estimation of the onset of the spinodal instability~\cite{Chaikin_Lubensky_1995}, and provides useful insights into the possible phases in this system. 

We show that for {\it{any}} choice of compositional disorder, the nonreciprocally interacting system is more stable than the passive one when one considers a system with infinitely many species with random interactions. The theoretical predictions are supported by the numerical solution of the full nonlinear model for a few chosen forms of the distribution for the average densities shown in Fig.~\ref{fig:Schematic}(b). We demonstrate the role of complex eigenvalues, focusing on finite number of components, in producing diverse dynamics that are not permitted in the equilibrium counterpart. The paper is organized as follows: we introduce the theoretical framework of the multi-species NRCH model in Section~\ref{sec:framework} motivating why it is essential to consider random Gaussian matrices whose diagonal matrices have been drawn from a different distribution as illustrated in Fig.~\ref{fig:Schematic} (a). In Section~\ref{sec:Spinodal}, we introduce a general condition pertaining to the linear stability of a system with compositional disorder.  In Section~\ref{sec:Spectrum}, we present the analytical arguments leading to the results in section~\ref{sec:Spinodal}. The varied dynamics possible in the model is explored in section~\ref{sec:StatisticsDynamics} and we present our conclusions in~\ref{sec:Conclusions}.

\section{multi-species NRCH with compositional disorder}\label{sec:framework}
\noindent
We follow the framework of \cite{parkavousi2024enhanced} for multicomponent active mixtures and consider $N$ active species, each conserved individually. All species interact via pairwise nonreciprocal interactions originating from the active nature of the system~\cite{tucci2024nonreciprocal,OuazanReboul2023}, and reciprocal interactions driven by a free energy. The scalar field $\phi_a({\bm r},t)$ represents the number density of the $a$-th species. The densities obey gradient dynamics of the form $\partial_t \phi_a = \Gamma_a \nabla^2 \mu_a$ with mobility $\Gamma_a$ and active chemical potentials $\mu_a$. The chemical potentials $\mu_a = \mu_a^{(\rm eq)}+\mu_a^{(\rm ac)}$ have two parts each: the first is the functional derivative of a free energy $F$, and the second represents the non-equilibrium active contribution that is not related to a free energy. The free energy functional $F = \int \mbox{d}^d r f$, where $f$ is the free energy density which promotes macroscopic phase separation. We use a minimal polynomial form for $f$ as shown below
\beq\label{eq:FE}
f &=& \sum_a \left[ \frac{1}{2} \Theta_a \phi_a^2 + \frac{s_a}{4} \phi_a^4 + \frac{K}{2} |\bm{\nabla} \phi_a|^2 \right] \nonumber \\
&& \frac{1}{2} \sum_{ab} \chi_{ab} \phi_a \phi_b.
\eeq
$\chi_{ab}= \chi_{ba}$ is Flory-Huggins type interaction parameter between two species $a$ and $b$. The active part of the chemical potential is
\beq 
\mu_{a}^{(\rm ac)} = \sum_{b} \alpha_{ab} \phi_b,
\eeq 
where $\alpha_{ab} = -\alpha_{ba}$.

To make progress, we have made some assumptions while still retaining enough complexity. The interfacial tension is assumed to be identical for all species and equal to $K$, in order to focus on effects that are distinct from a Turing type instability that appears when the coefficients are chosen to be distinct~\cite{ThieleTuring}. The coefficient of self-interaction $\Theta_a = (1+\Theta)$ is the same for all species, a parameter akin to an external tuning parameter, such as the temperature of the system, that controls the transition to a patterned state. Rescaling time, space, and the scalar fields as $t \to t {K}/{\Gamma}$, $x \to x\sqrt{K}$, and $\phi_a \to \phi_a/\sqrt{s_a}$, the equations can be recast as follows by introducing an interaction matrix $\mathbb M$ with elements $M_{ab} = \chi_{ab} + \alpha_{ab}$
\beq\label{eq:multiNRCH}
&& \partial_t \phi_a = \nabla^2   \left[ (1+\Theta) \phi_a + \phi_a^3 + \sum_{b} M_{a b} \phi_b \right] \nonumber \\
&& - \nabla^4 \phi_a, \,\, \int  \mbox{d}^d \br \phi_a(\br ,t) = \phi_{a0},
\eeq
where the term quartic in space gradients represents the surface tension and appears due to the interfacial contribution to $f$ in Eq.~\eqref{eq:FE}. From this point onwards, we consider an ensemble of systems where all elements of $\mathbb{M}$ are drawn from a Gaussian distribution of zero mean and variance $\sigma$, a quantity that sets the scale of the inter-species interactions.
\begin{equation}\label{eq:InteractionMatrix}
M_{ab}(\bar{\chi}, \bar{\alpha})=\frac{\bar{\alpha}}{2\sqrt{N (\bar{\alpha}^2+\bar{\chi}^2)}} A_{ab}+\frac{\bar{\chi}}{2\sqrt{ N(\bar{\alpha}^2+\bar{\chi}^2)}} S_{ab},
\end{equation}
where $\mathbb{A}$ and $\mathbb{S}$ are fully asymmetric and symmetric matrices respectively, i.e. $A_{ab} = -A_{ba}$, and $S_{ab} = S_{ba}$.  Eq.~\eqref{eq:InteractionMatrix}  follows from the general rule that any matrix can be expressed as the sum of a symmetric and an antisymmetric matrix. Eq.~\eqref{eq:InteractionMatrix} is written such that the variance of the coefficients is $\sigma^2$ and the correlation between elements $M_{ab}$ and $M_{ba}$ of $\mathbb{M}$ determines the level of nonreciprocity parameterized by $\alpha$ which depends on the ratio of $\bar{\alpha}$ and $\bar{\chi}$. The properties of $\mathbb{M}$ are summarized as follows
\beq\label{eq:LinStable}
&&\langle M_{ab} M_{ab} \rangle = \frac{\sigma^2}{4N}, \,\, \langle M_{aa} \rangle = 0 , \nonumber \\
&& \langle M_{ab} M_{ba} \rangle = \frac{\sigma^2}{4N}(1 - 2\alpha^2), \,\, \alpha = \frac{\bar{\alpha} \bar{\chi}^{-1}}{\sqrt{\bar{\alpha}^2 \bar{\chi}^{-2} + 1}}.
\eeq
where the angular brackets denote averaging over ensembles of $\mathbb{M}$.

To determine the linear stability of the system we substitute $\phi_a = \phi_{a0} + \delta \phi_a$ in Eqs.~\eqref{eq:multiNRCH} to obtain the following linearised dynamics in Fourier space with wavenumber $q$ retaining terms only quadratic order in $q$
\beq \label{eq:LinDyn}
&& \partial_t \delta \phi_a \nonumber \\
&& = -q^2 \sum_b \left[  M_{ab} + (1 + \Theta + 3 \phi^2_{a0}) \delta_{ab}  \right] \delta \phi_b + \mbox{O}(q^4). 
\eeq 
Notice that the parameter $\sigma$ simply renormalizes the effective temperature and can be set to unity without any lack of generalization in all calculations, other than in Section~\ref{sec:Spectrum} where we discuss the threshold of stability using analytical arguments. Defining a diagonal matrix $C_{aa} = 3 \phi^2_{a0}$, whose off-diagonal entries are zero, it is clear from Eq.~\eqref{eq:LinDyn} that the linear stability of the system is determined by the eigenvalues of the matrix $\mathbb{D} = \mathbb{M} + \mathbb{C}$. The effect of varied composition thus enters the dynamical matrix as diagonal disorder through the diagonal matrix $\mathbb{C}$ whose diagonal elements $\bar{\phi }$ are drawn from a distribution $P(\bar{\phi})$, as illustrated in Fig.~\ref{fig:Schematic} (a). 

\begin{figure*}
	\centering
	\includegraphics[width= 0.99\linewidth]{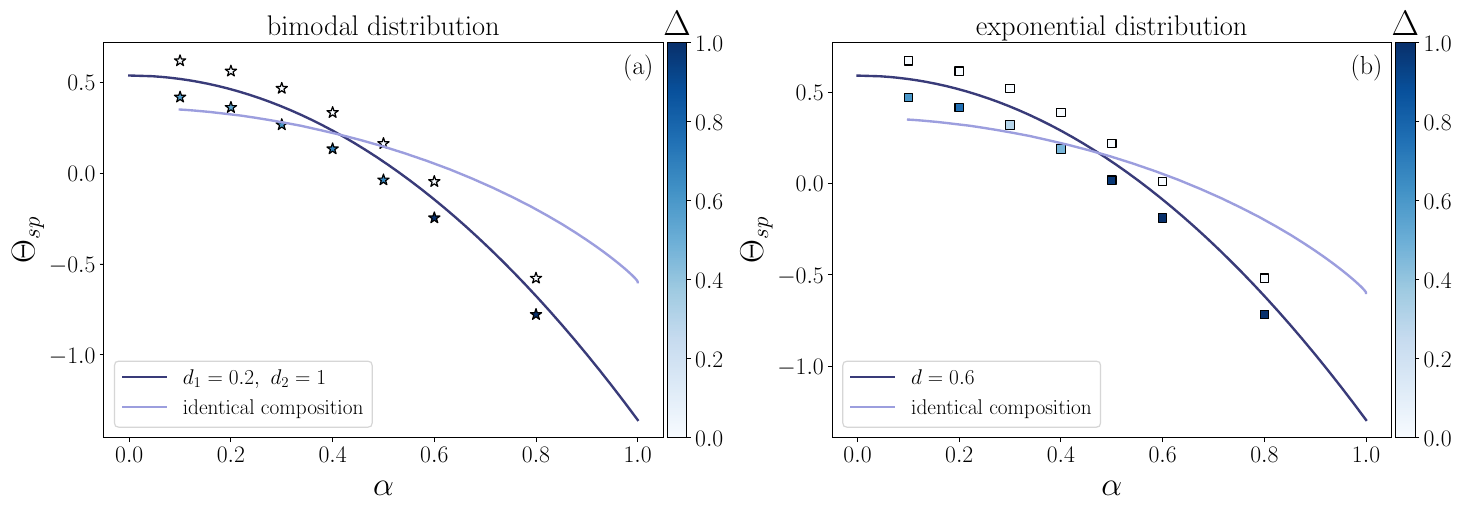}
	\caption{\textbf{Spinodal instability in theory and simulations}: (a) The boundary of linear stability is plotted with a solid line in the $\Theta - \alpha$ plane, for a fixed set of parameters for the discrete bimodal distribution, see Eq.~\eqref{eq:Bimodal}. The predictions are tested using simulations with $50$ realisations of $\mathbb{M}$. The colour of the markers reflects the ensemble and space-averaged order parameter $\Delta$ [see Eq.~\eqref{eq:OrderParameter}], which vanishes outside the spinodal shown by a light hue and changes discontinuously to a finite value inside the unstable region, as marked by a darker hue. The grey line is the same for an identical mean composition of the densities. The plots show that the relative stability depends both on the distribution and on $\alpha$  (b) Same as panel (a) for the exponential distribution. } 
	\label{fig:Simulations1}
\end{figure*}

\section{Spinodal}\label{sec:Spinodal}
\noindent
The stability of the system has to be determined in a multidimensional space constituted by the the effective temperature $\Theta$ and the parameters defining the distribution $P$ from which the average densities of the species are selected. The boundary of the spinodal instability separates the part of the multi-dimensional parameter space  where at least one of the N eigenmode in Eq.~\eqref{eq:LinStable} is linearly unstable (and grows exponentially) from the region where the homogeneously mixed state is stable. The onset of the spinodal instability is thus determined by the eigenvalue $\lambda_0$ of $\mathbb{D}$ with the smallest real part. With this definition, we can discuss the onset of pattern formation in the system as controlled by the effective temperature $\Theta$.

Setting the R.H.S. of Eq.~\eqref{eq:LinDyn} to zero, we determine the the threshold value of $\Theta = \Theta_{\rm sp}$ for the mixed state to be stable to perturbations.
\beq \label{eq:Spinodal}
\Theta_{\text {sp }}\left(\alpha, \phi_0, \{D\}\right)=-1-\operatorname{Re}\left[\lambda_0\left(\alpha, \{D\}\right)\right],
\eeq 
where $\{ D \} $ denotes the set of parameters that characterize the distribution $P$. On lowering $\Theta$ below $\Theta_{\rm sp}$, the mixture undergoes a spinodal instability as shown in the kymographs of Fig.~\ref{fig:Schematic} (c) to evolve to a patterned final state. The statistics of $\lambda_0$ are determined by the properties of the dynamical matrix $\mathbb D$. For $\Theta$ lower than $\Theta_{\rm sp}$, the homogeneous state is unstable and patterns are formed in the steady state. 

For $P(\bar{\phi}) = \delta(\bar{\phi} - \phi_0)$
, i.e. if the mean density of each species is identical, $\lambda_0$ is given by the eigenvalues of $\mathbb{M}$ shifted by a constant factor~\cite{parkavousi2024enhanced}. 
\beq \label{eq:ThetaSp}
\Theta_{\rm sp} = - 3 \phi_0^2 -\alpha^2.
\eeq
Eq.~\eqref{eq:ThetaSp} implies that for a fixed $\phi_0$, $\Theta_{\rm sp}$ is smaller for the active system in comparison to the passive system ($\alpha = 0$). The control parameter $\Theta + 1$ is synonymous with the strength of self-interaction. If all entries of the matrix $\mathbb{M}$ were set to zero, the non-interacting system would not phase separate for $\Theta>-1$. $\Theta<-1$ implies that interactions between particles of the similar is attractive and leads to phase separation. Eq.~\eqref{eq:ThetaSp} shows that random passive interactions lead to $\Theta_{\rm sp} = 0$, meaning that they generate effective attractive interactions between the species, such that to tune the system to a homogeneous state, it is necessary to increase intra-species repulsive interactions. For a fully nonreciprocal system, $\alpha = 1$, $\Theta_{\rm sp}$ decreases to $-1$ which is the threshold for the non-interacting system. The nonreciprocal counterpart is thus more stable than the reciprocal system. The factor of unity in Eq.~\eqref{eq:Spinodal} simply sets $\Theta_{\rm sp}$ for passive interactions to zero which serves as a good reference point for comparisons. 

We will now state the first important result of our work before illustrating its significance for specific choices of $P$. We find that that for any choice of $P$, $\Theta_{\rm sp}$ satisfies the following simple relation
\beq \label{eq:slope}
&& \Theta_{\rm sp} = -1 -\mbox{Re}[\lambda_0] = l_0 + m\alpha^2.
\eeq 
The slope $m$ and $l_0$ depend only on the parameters ${D}$ characterizing the distribution $P$; in particular, both of these parameters are independent of $\alpha$. The relation in Eq.~\eqref{eq:slope} is rigorously true in two cases; it is a good estimate for an ensemble of systems with interactions following the statistical rules laid out in Eq.~\eqref{eq:InteractionMatrix} for finite $N$, and Eq.~\eqref{eq:slope} is rigorously true for a system with infinitely many components. When $N \to \infty$, the variance of $\mbox{Re}(\lambda_0)$, which goes down with $N$ as $1/\sqrt{N}$, approaches zero. For finite $N$ we expect deviations from the relation in Eq.~\eqref{eq:slope}. Keeping this in mind, we choose $N = 100$ in most of the simulations that are carried out to verify Eq.~\eqref{eq:slope}, as for this value we expect fluctuations of $10\%$ in $\lambda_0$ for $\alpha = 0$, which converges to $-1$ for $N \to \infty$. The numerical results presented in the paper are also averaged over an ensemble of systems to compensate for the expected variations.

First consider the system with only reciprocal interactions ($\alpha = 0$), with any specified $P$. Notice that $l_0$ is the only parameter that determines the stability of the passive system in the presence of compositional disorder. For different distributions $P$, we can compute $l_0$ and use the results to compare the effect that a particular choice of $P$ has on the stability. The lower the value of $l_0$, the more stable the system is, as the transition to a patterned state happens at a lower effective temperature (see discussion after Eq. 8). We now discuss the stability of the uniform mixture on tuning the level of nonreciprocity $\alpha$. The sign of the `slope' $m$ between $\Theta_{\rm sp}$ and $\alpha^2$ determines the stability of the active system. For $m<0$ the active system is more stable than the passive one. We will show in Section~IV that the slope $m$ is always negative, irrespective of the details of the distribution $P$. This implies a very general consequence of nonreciprocity is the stabilisation of the mixed state. We will also show that the sign of $l_0$ depends on the specific choice of $P$.

In Fig.~\ref{fig:Simulations1} we have plotted the theoretical prediction for $\Theta_{\rm sp}$ for the bimodal and exponential distributions described in \ref{sec:Spectrum}, see Eqs.~\eqref{eq:Bimodal} and \eqref{eq:expo}. The solid lines in Fig.~\ref{fig:Simulations1} separate the region where the uniform reference is stable (higher $\Theta_{\rm sp}$) from the unstable region (higher $\Theta_{\rm sp}$). To test the predictions of linear stability analysis, we solve Eq.~\eqref{eq:multiNRCH} numerically (see Appendix~\ref{App:Numerics} for details). At the beginning of the time evolution, the scalar fields are uniform with mean $\phi_{a0}$ such that $3 \phi_{a0}^2$ is drawn from either the bimodal or the exponential distributions. The initial conditions at the beginning of the numerical simulation is thus $\phi_a = \phi_{a0} + \delta \phi_a$, where small deviations $\delta \phi_a$ are uncorrelated in space and drawn from a Gaussian distribution. For other details on system size, methods used, see the Appendix~\ref{App:Numerics}. The results of the simulation are shown in Fig.~\ref{fig:Simulations1}, where the markers denote the points where the simulations are carried  out and the colour of the markers carries information about the magnitude of the order parameter.
\beq\label{eq:OrderParameter}
\Delta =  \int \mbox{d}x  \frac{1}{N}\sum_a \left(\phi_a(x,t)- \phi_{a0} \right)^2.
\eeq
$\Delta$ measures the mean squared deviation from the homogeneous state and vanishes when it is stable. $\Delta$ deviates from zero when the system develops patterns, as seen in the space-time kymograph showing the evolution of the system in one dimension in Fig.~\ref{fig:Schematic} (c). Thus, $\Delta$ serves as the order parameter for the transition to the ordered state. The value of $\Delta$ is determined by a spatial and temporal averaging of the numerical results averaging, followed by an ensemble averaging over $50$ realizations of the matrix $\mathbb{M}$, for $N=100$ components. The agreement between theoretical and numerical results in Fig.~\ref{fig:Simulations1} for both bimodal and exponential distributions (panels (a) and (b)) proves that the stability condition in Eq.~\eqref{eq:slope} which holds for infinitely many components is applicable for a finite number of components as well.

\begin{figure}
	\centering
	\includegraphics[width= 0.99\linewidth]{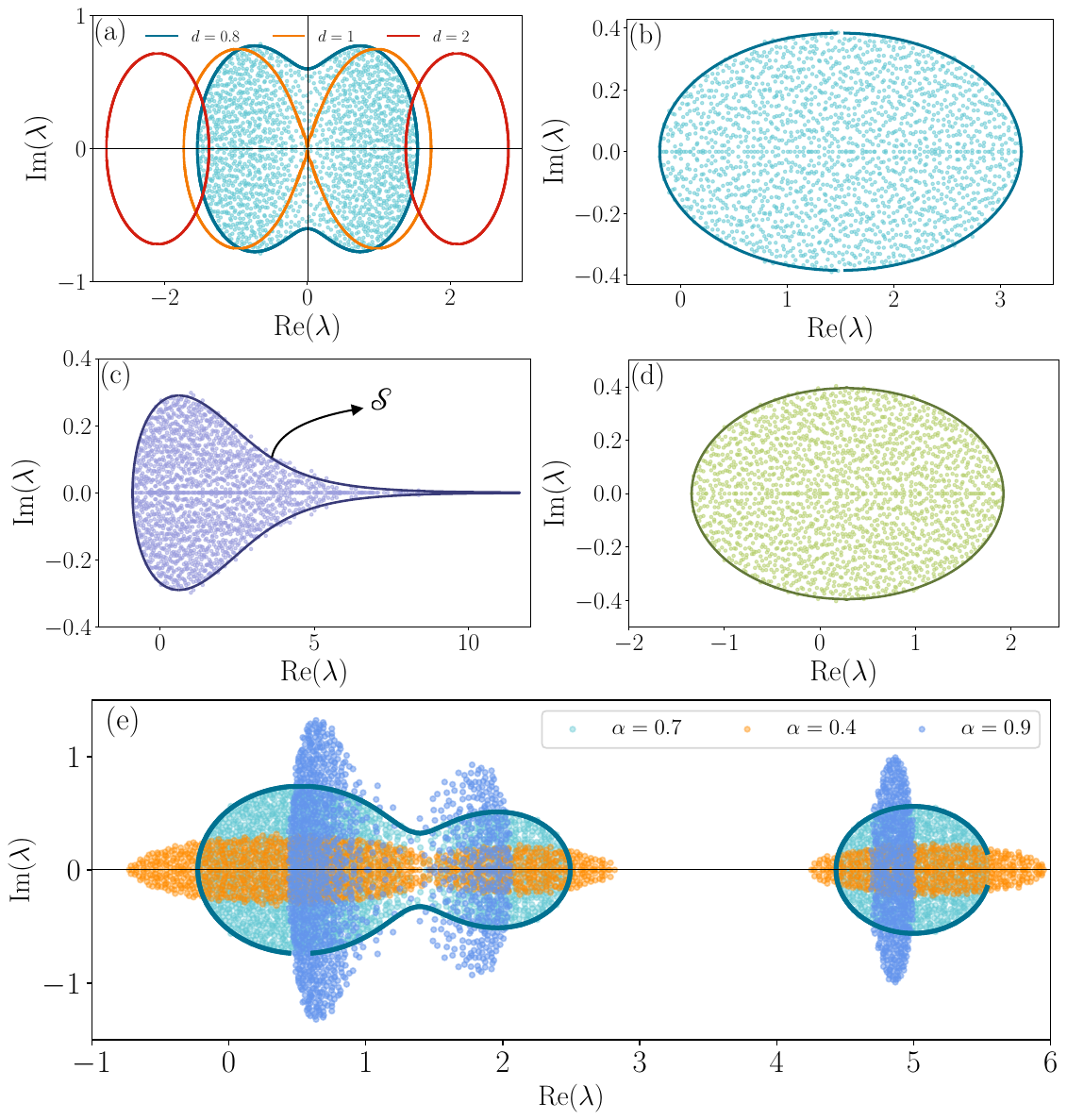}
	\caption{\textbf{Boundary enclosing the eigenvalues in the complex plane}: (a) The boundary and the spectrum of a large random matrix with diagonal disorder drawn from the bimodal distribution, see Eq.~\eqref{eq:Bimodal} for $d_2 = -d_1 = d$. {(b)} The diagonal elements of $\mathbb{C}$ are drawn from a uniform distribution where $d_{-}=1$ and $d_+=2$. {(c)} The diagonal disorder is drawn from an exponential distribution with $d=0.6$. {(d)} The spectrum and the boundary of support for a large random matrix with diagonal elements from a beta distribution with $a=2$ and $b=5$. For all examples, the random matrix is of the size $\text{N}=1000$ and $\chi=1,\alpha=0.5$. {(e)} The diagonal disorder are drawn from a trimodal distribution with $d_1= 0.5$, $d_2= 2$, $d_3=5 $,  $p_1= {1}/{2}$ and $p_2={1}/{5} $ the random matrix is of a size $N=2000$ and $\chi=1$. }
	\label{fig:Boundary}
\end{figure}

\begin{figure*}
	\centering
	\includegraphics[width= 0.99\linewidth]{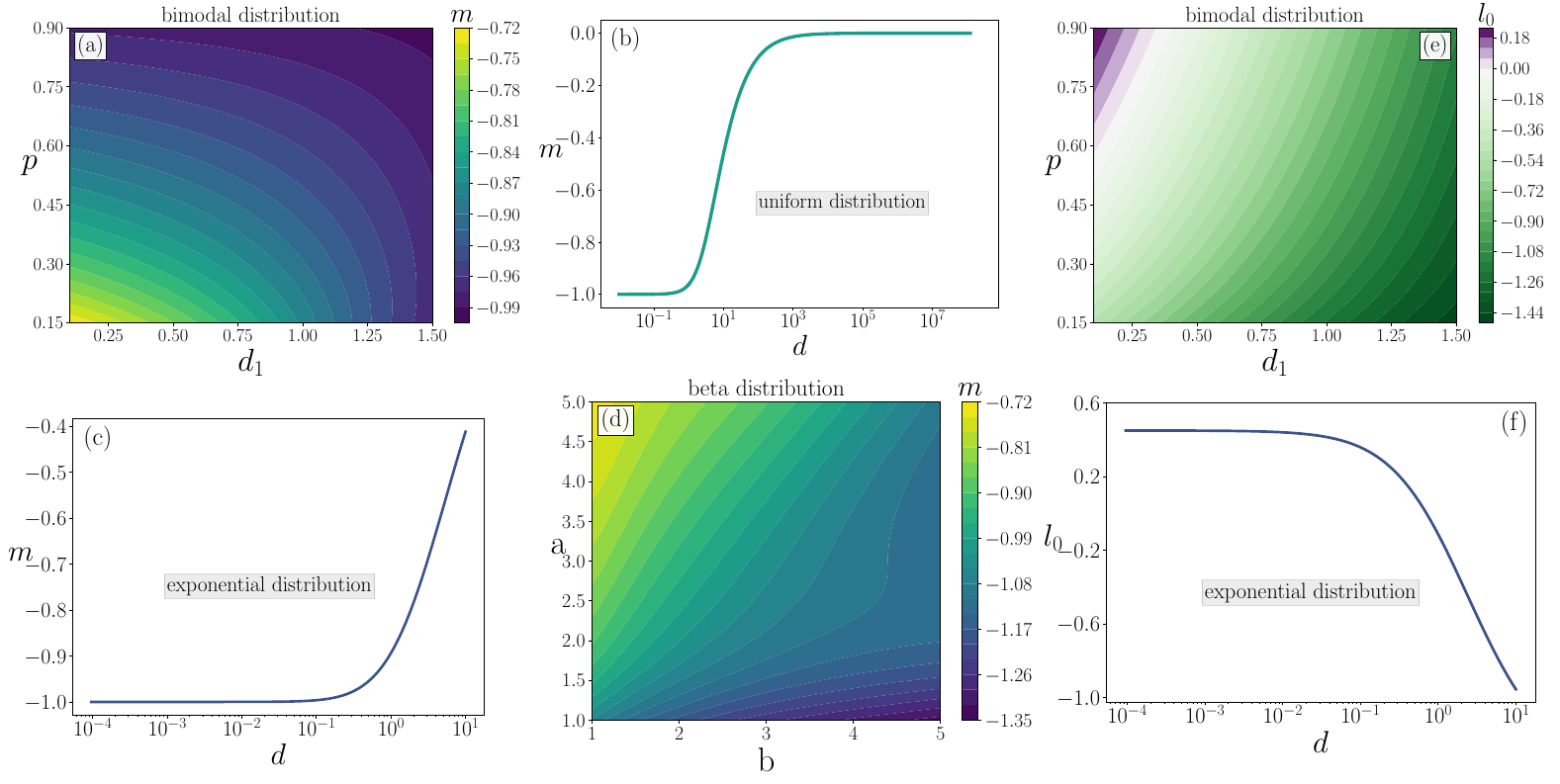}
	\caption{\textbf{The slope $m$ and the intercept $l_0$ for different types of compositional disorder}: The slope $m$, referred to in Eq.~\eqref{eq:slope} and \eqref{eq:FinalSlope} for the definition, is calculated for the distributions indicated in the figure. The heatmap of $m$ for the two parameters in (a) and (d) and its variation with the width $d$ in (b) and (c) shows that it is always less than zero. Both the uniform and exponential distributions approach the delta function for $d \to 0$ so that $m$ approaches $-1$ in panels (b) and (c). Panels (e) and (f) show that $l_0$ can be positive or negative, meaning that $\Theta_{\rm sp}$ for the passive system can be higher or lower for the corresponding system with identical mean densities.}
	\label{fig:Slopes}
\end{figure*}
\section{Eigenspectrum and the most unstable eigenvalue}\label{sec:Spectrum}
\noindent
The objective of this section is to establish the results in the previous section with analytical arguments. We first explore the eigenvalue spectrum due to compositions drawn from four different distributions that are qualitatively different from one another, being either bounded or unbounded, and which assume either discrete or continuous values. The boundary enclosing the spectrum of complex eigenvalues is approximated by a continuous curve when $N \to \infty$, as illustrated in Fig.~\ref{fig:Boundary} with some examples. In Fig.~\ref{fig:Boundary}, the markers denote eigenvalues of a single realization of the matrix $\mathbb{D}$ with $N=2000$, while the boundary has been calculated using methods described in \cite{cure2023antagonistic} which provides an estimate for an infinite number of components. We will briefly describe the analytical approach and use it to justify the form of Eq.~\eqref{eq:slope}. The spectral distribution $\rho$ of the eigenvalues $\lambda_j$ of $\mathbb{D}$ can be written formally as
\begin{equation}
\rho(z)=\lim _{n \rightarrow \infty} \frac{1}{n}\left\langle\sum_{j=1}^n \delta\left(x-\operatorname{Re}\left(\lambda_j\right)\right) \delta\left(y-\operatorname{Im}\left(\lambda_j\right)\right)\right\rangle,
\end{equation}
The boundary of the eigenspectrum of $\mathbb{D}$ (one that encompasses all eigenvalues $\lambda_j$ in the complex plane) or equivalently, the support set of $\rho(z)$ is called $\mathcal{S}$ and is given by the points $z \equiv (x,y)$ in the complex plane as indicated in Fig.~\ref{fig:Boundary}.

As the name suggests, the discrete bimodal distribution is a discrete distribution in a bounded domain with delta peaks at two values $d_1$ and $d_2$.
\beq \label{eq:Bimodal}
P(\bar{\phi}) = p \delta( \bar{\phi}-d_1) + (1-p) \delta(\bar{\phi}-d_2).
\eeq 
As seen in Fig.~\ref{fig:Boundary} (a), the bimodal effectively splits the components into two groups containing $pN$ and $(1-p)N$ species each. For any $p$ and for $d_{1,2} \to 0$ the boundary of the spectrum is simply stretched along the real axis. As $d_{1,2}$ increases, a fissure appears in the boundary, whose location reflects the chosen $p$, being located around $\mbox{Re}(\lambda) = 0$ for $p=1/2$. Eventually, at a large enough value of $d$, $\mathcal{S}$ splits into two dis-connected contours, each enclosing $pN$ and $(1-p)N$ eigenvalues. The numerical results match well with the contour determined from solving Eqs.~\eqref{eq:Boundary1} and \eqref{eq:Boundary2}, as seen in Fig.~\ref{fig:Boundary} (a). The distribution can be generalised for multiple discrete values $d_i$ with probability $p_i$ as follows
\beq \label{eq:BimodalMany}
P(\bar{\phi}) = \sum_{i=1}^{n} p_i \delta( \bar{\phi}-d_i), \,\, p_n = 1 - \sum_{i=1}^{n-1}p_i.
\eeq 
An example of the trimodal distribution and a splitting of the spectrum into three parts in shown in Fig.~\ref{fig:Boundary} (e).

The exponential distribution is unbounded and described by a single parameter $d$ that determines the upper-limit of the values that are sampled with high probability.
\beq \label{eq:expo}
 P(\bar{\phi}) = \frac{1}{d} \exp\left( -\frac{\bar{\phi}}{d} \right), \,\, d>0.
\eeq 
The boundary in Fig.~\ref{fig:Boundary} (b) reflects the asymmetry of the distribution and shows a tail that reflects the large deviations, as only a few species are included with high mean composition.

The uniform distribution is defined by its width $d$ within which all compositions are sampled with the same probability
\beq \label{eq:uniform}
P(\bar{\phi}) =  \frac{1}{d_{2} - d_{1}}, \,\, \bar{\phi} \in [d_{1},d_{2}]  ,
\eeq 
shown in Fig.~\ref{fig:Boundary} (c) shows an elongated boundary along the real axis. With increasing $d$, the aspect ratio increases as well. 

Lastly, we examine the Beta distribution, which is a continuous and bounded distribution parameterized by indices $a$, $b$, and its bound $d$.
\beq \label{eq:beta}
&& P(\bar{\phi})=N_{a, b}^{-1}\ \bar{\phi}^a \left(d -  \bar{\phi} \right)^b, \quad x \in\left[0, d\right] \nonumber \\
&& N_{a, b}=d^{a+b+1} B(a+1, b+1),
\eeq
where $N_{a,b}$ is the normalization constant and $B$ is the Beta function. The method used to calculate the boundary $\mathcal{S}$ is described next.
\\\\
\subsection{Calculation of the boundary $\mathcal{S}$}
\noindent
The starting point of our analysis are the coupled equations, see Eqs.~\eqref{eq:Boundary2} and \eqref{eq:Boundary1} below, that determine the contour of points $(x,y)$ in the complex plane enclosing the eigenspectrum $\{\lambda_j\}$ of $\mathbb{D}$.  To do so, it is essential to introduce the function $\mbox{Re}(g)$ which is related to the resolvent $\sum_{j} (z - \lambda_{j})^{-1}$~(see Appendix~\ref{App:Boundary} for details and \cite{cure2023antagonistic} for the original derivation). The condition that $\mbox{Re}(g)$ assumes finite values outside the boundary $\mathcal{S}$ (i.e. in the region where no eigenvalues appear) and diverges precisely on the boundary was used to determine the relation between the points $(x,y)$ constituting $\mathcal{S}$ and $\mbox{Re}(g)$~\cite{cure2023antagonistic}. The necessity of introducing $\mbox{Re}(g)$ has been explained in the Appendix~\ref{App:Boundary}.
\beq
&& \int \mbox{d}\bar{\phi}  P(\bar{\phi})  \, \frac{\sigma^2}{\left(( \bar{\phi} -x)+\operatorname{Re}\left(g\right) (1-2 \alpha^2) \sigma^2\right)^2+{y^2}/{(2\alpha^2)^2}} \nonumber \\
&& = 1.
\label{eq:Boundary2}
\eeq
and
\begin{eqnarray}
&& \int \mbox{d}\bar{\phi} P(\bar{\phi})\, \frac{\bar{\phi}-x+\operatorname{Re}\left(g\right) (1-2 \alpha^2) \sigma^2}{\left((\bar{\phi}-x)+\operatorname{Re}\left(g\right) (1-2 \alpha^2) \sigma^2\right)^2+ {y^2}/{(2\alpha^2)^2}} \nonumber \\
&& = -\operatorname{Re}\left(g\right),
\label{eq:Boundary1}
\end{eqnarray}
where the integrals are defined over the domain of $\bar{\phi}$. For $P(\bar{\phi}) = \delta(\bar{\phi} - \phi_0)$, the eigenvalues are distributed uniformly with an ellipse with major and minor axes equal to $1-\alpha^2$ and $\alpha^2$ respectively, such that $\mbox{Re}[\lambda_0] = 1-\alpha^2$ as verified readily from Eqs.~\eqref{eq:Boundary1} and \eqref{eq:Boundary2}, see Appendix~\ref{App:Identical}.

The general relation in Eq.~\eqref{eq:slope} is motivated by a close examination of the structure of the coupled equations Eqs.~\eqref{eq:Boundary1} and \eqref{eq:Boundary2} and verified by numerical calculations using the distributions introduced above. To calculate $\mathcal{S}$ for any distribution $P$, we define scaled and shifted coordinates $\tilde{x}$ and $\tilde{y}$
\beq \label{eq:redef}
\tilde{x} = x - \mbox{Re}(g)(1-2 \alpha^2)\sigma^2, \, \, \tilde{y} = \frac{y}{ 2\alpha^2},
\eeq
Notice that Eqs.~\eqref{eq:Boundary2} and \eqref{eq:Boundary1} rewritten using $\tilde{x}$ and $\tilde{y}$ are independent of $\alpha$, a property that will be used in evaluating $\mathcal{S}$. As a first step to calculating $\mathcal{S}$ we use Eq.~\eqref{eq:redef} to re-write Eq.~\eqref{eq:Boundary2} as
\beq \label{eq:Rel2}
\frac{1}{\sigma^2} = \int \mbox{d} \bar{\phi} \, P(\bar{\phi}) \frac{1}{(\bar{\phi}- \tilde x)^2 + \tilde{y}^2} \equiv  h({\tilde{x}}; \tilde{y}),
\eeq 
thus defining the function $h_{\tilde y}$. The function $h_{\tilde y}$ can in principle be inverted to obtain $\tilde x$, for a fixed $\tilde y$ as 
\beq 
\tilde x = h^{-1}(\sigma^{-2};{\tilde y}).
\eeq 
In the second step, $\mbox{Re}(g)$ is calculated using Eq.~\eqref{eq:Boundary1},
\beq \label{eq:reG}
\mbox{Re}(g)[\tilde{x},\tilde{y}]= -\int \mbox{d}\bar{\phi} P(\bar{\phi})\, \frac{\bar{\phi}-\tilde{x}}{(\bar{\phi}-\tilde{x})^2+ {\tilde{y}^2}}.
\eeq 
The following expression for $x$ holds for a given $y$
\beq \label{eq:Xexpression}
 && x = h^{-1}(\sigma^{-2},{\tilde y}) \nonumber \\
 && + (1-2 \alpha^2) \sigma^2 \mbox{Re}(g)[ h^{-1}(\sigma^{-2};{\tilde y}), \tilde{y}],
\eeq 
showing that $x$ depends explicitly on $\alpha$ only through the second term in the R.H.S. of Eq.~\eqref{eq:Xexpression}. The procedure yields the set of points $(x,y)$ defining the curve $\mathcal{S}$ that serves well as the boundary for the eigenvalues plotted for $N=2000$ in the complex plane, see Fig.~\ref{fig:Boundary}.

$\alpha = 1/\sqrt{2}$ is a special value of $\alpha$ because at this point $\tilde{x} = x$, and it is particularly easy to carry out the algebraic operations described above, a point illustrated with the example of the bimodal distribution in Appendix~\ref{App:Boundary}. The physical meaning of $\tilde x$ and $\tilde y$ is now clear - they are simply the curve $\mathcal{S}$ at $\alpha = 1/\sqrt{2}$. Consider now two values of $\alpha$, $\alpha_{1,2}$ both different from $1/\sqrt{2}$, at fixed $y$ we have using Eq.~\eqref{eq:redef}   
\beq \label{eq:curveS}
&& \frac{x(\alpha_1) - x(\alpha_2)}{\alpha_2^2 - \alpha_1^2} = 2 \left[\mbox{Re}(g)(\tilde{x},\tilde{y}) \right] \sigma^2 \nonumber \\
&& \frac{y(\alpha_1)}{y(\alpha_2)} = \frac{\alpha_2^2}{\alpha_1^2}.
\eeq
Eq.~\eqref{eq:curveS} suggests that the curve $\mathcal{S}$ can be calculated at all values of $\alpha$ if we calculate it at $\alpha = 1/\sqrt{2}$. 

Notice now that for two different values of $\alpha$, the $y$ values of the curves $\mathcal{S}$ are proportional. For $\tilde{y} = 0$, i.e., the points at which $\mathcal{S}$ intersects the $x$ axis are the same for $\mathcal{S}$ for any value of $\alpha$. This implies that the topology of $\mathcal{S}$, meaning the number of times the curve $\mathcal{S}$ intersects the real axis, is determined by $\{D\}$ only and does not depend on $\alpha$ at all, as illustrated using the bimodal distribution in panel (e) of Fig.~\ref{fig:Boundary}. The curve in $\mathcal{S}$ splits into two above a threshold value of $d$, as seen in Fig.~\ref{fig:Boundary} (a), meaning that $\tilde{y}$ vanishes at four points on the $x-$axis. The topology of the boundary is preserved for all values of $\alpha$ as shown in Fig.~\ref{fig:Boundary} (e) for a trimodal distribution where the $\{ D \}$ are fixed and $\alpha$ only is varied. The spectrum for the values indicated at the values of alpha indeed follows the insight from theory.

\subsection{Expression for $\lambda_0$}
\noindent
By the definition of $\lambda_0$ (intercept $x$ of the curve $\mathcal{S}$ at $y=0$), we have using Eq.~\eqref{eq:Xexpression}
\beq \label{eq:FinalSlope}
\lambda_0(\alpha) = L_0 - m \alpha^2,
\eeq
where the intercept $L_0$ that determines the threshold of instability for the passive system is given by 
\beq \label{eq:Intercept}
&& L_0 = h^{-1}\left( \sigma^{-2};0 \right) +  \sigma^2 \mbox{Re}(g)[h^{-1} \left(\sigma^{-2}; 0 \right),0].
\eeq
$\lambda_0$ varies quadratically with $\alpha^2$ with the constant of proportionality or 'slope' $m$ given by
\beq \label{eq:m}
m = 2 \sigma^2 \mbox{Re}(g)[h_{0}^{-1}\left(-\sigma^{-2} \right),0] .
\eeq
Substituting Eq.~\eqref{eq:m} in the definition of the threshold of instability $\Theta_{\rm{sp}} = -1 - \mbox{Re}[\lambda_0]$ produces the expression in Eq.~\eqref{eq:slope} with the identification $l_0 = - 1-L_0$.

 \begin{figure}
	\centering
	\includegraphics[width= 0.99\linewidth]{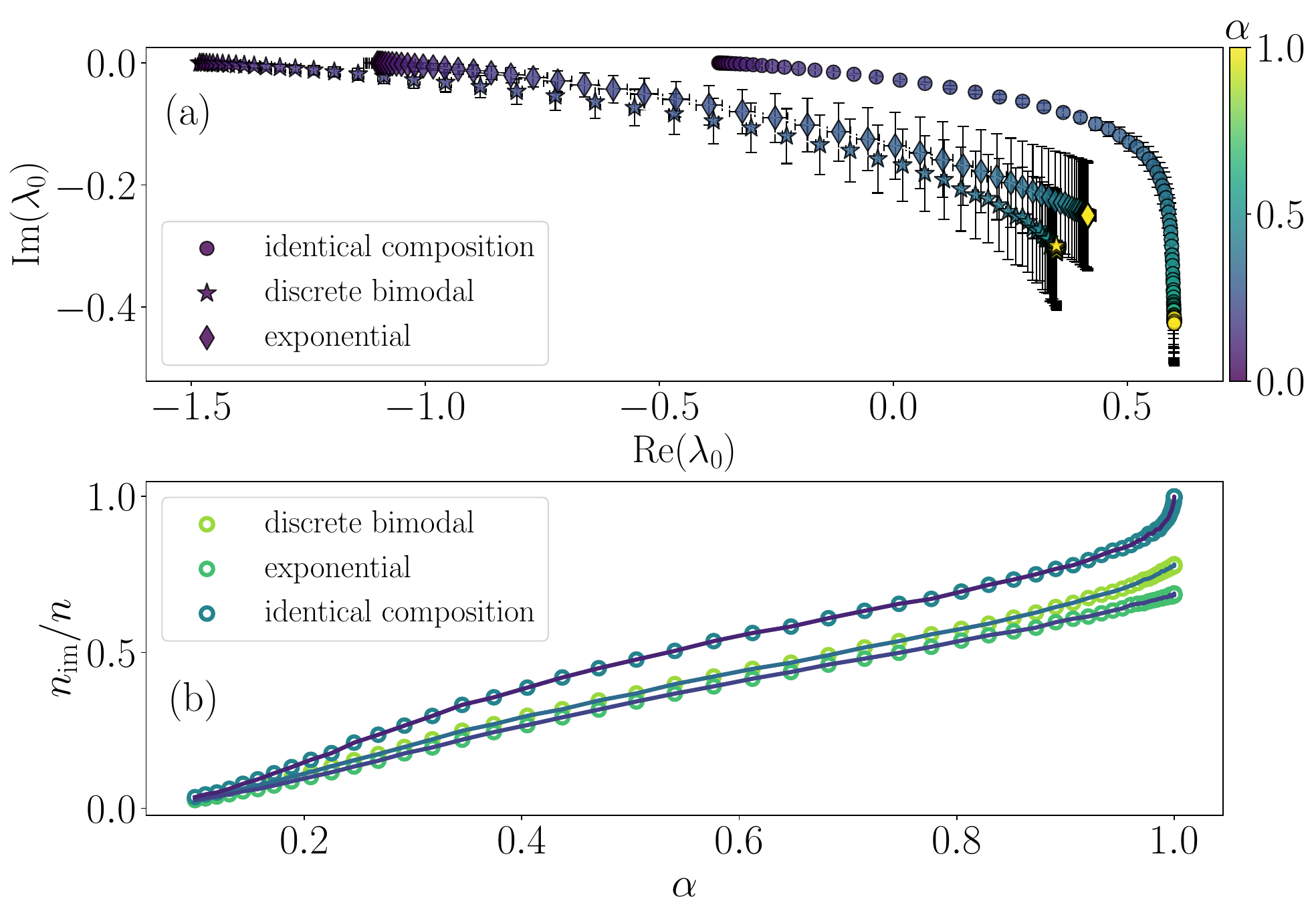}
	\caption{\textbf{Statistics of $\lambda_0.$} In the limiting case of infinite conserved densities, $\lambda_0$ is real. For any finite-sized system, $\langle \lambda_0 \rangle$ averaged over ensembles of $\mathbb{D}$, is in fact complex. (a) The imaginary and real parts, $\mbox{Im}(\lambda_0)$ and $\mbox{Re}(\lambda_0)$, are plotted as a function of $\alpha$, encoded in the colour of the marker. The error bar indicates the variance. At each value of $\alpha$, we consider $10^4$ realizations of $\mathbb{D}$ for $N=300$. From the plots, it is clear that the variability in $\mbox{Im}(\lambda_0)$ depends on the type of compositional disorder. Notice also that the variations are increased compared to the uniform average composition. (b) The fraction of complex eigenvalues is plotted as a function of $\alpha$, showing that the fraction increases with $\alpha$ although it does not approach unity. }
	\label{fig:Stats}
\end{figure}
From Eq.~\eqref{eq:m}, it is evident that $m$ depends only on the parameters $\{ D\}$ meaning that it is enough to resolve the sign of $m$ for any non-zero value of $\alpha$, which following the train of discussion so far in the section is accomplished easily at $\alpha = 1/\sqrt{2}$. 

The form of the diagonal terms that we consider represents squared deviations of the mean density of each species from the average and thus they are always positive quantities bounded by zero from below. From Eq.~\eqref{eq:redef} it is clear that at the extreme value $\lambda_0$ of $\tilde{x} = x$, 
\beq 
m = -2 \sigma^2 \int \mbox{d}\bar{\phi} P(\bar{\phi})\, \frac{1}{(\bar{\phi}-\lambda_0(1/\sqrt{2}))} 
\eeq 
the sign of the integrand in Eq.~\eqref{eq:reG} is determined by the sign of $(\bar{\phi}-\lambda_0(1/\sqrt{2}))$, a quantity whose sign cannot change in the domain of $\bar{\phi}$ for a distribution with a well-defined $\lambda_0$. This argument constraints the signs of the deviation $\bar{\phi} - \lambda_0({1/\sqrt{2}})$ and thus $m$ to be either always negative and positive respectively, or vice versa. The two cases represent the two extreme bounds of the eigenvalues, and clearly $m>0$ corresponds to the eigenvalue with the smallest real part.

To test the validity of the theoretical predictions, we plot $m$ by varying the parameters $\{ D \}$ for the distribution shown in Fig.~\ref{fig:Schematic}. For each value indicated in the axes of the panels in Fig.~\ref{fig:Slopes}, we calculate $\lambda_0$ at two values of $\alpha^2$ and calculate the slope and verify that it is indeed positive everywhere (see panels (a-d) of Fig.~\ref{fig:Slopes}).
\beq \label{eq:ExpSlopes}
m = \frac{\lambda_{0}(\alpha_1) - \lambda_{0}(\alpha_2)}{\alpha^{2}_2 - \alpha^{2}_1}.
\eeq
We also calculate $L_0$ and find that it can assume both positive and negative signs depending on the parameters specifying the distribution meaning that the stability of the passive system, ($\alpha = 0$) does not follow a universal trend. Panels (e) and (f) of Fig.~\ref{fig:Slopes} shows the variation in $\Theta_{\rm{sp}}(0) = l_0 = -L_0-1$, (refer to Eq.~\eqref{eq:slope}) for the bimodal and exponential distributions respectively.  

\begin{figure*} 
	\centering 
	\includegraphics[width= 0.99\linewidth]{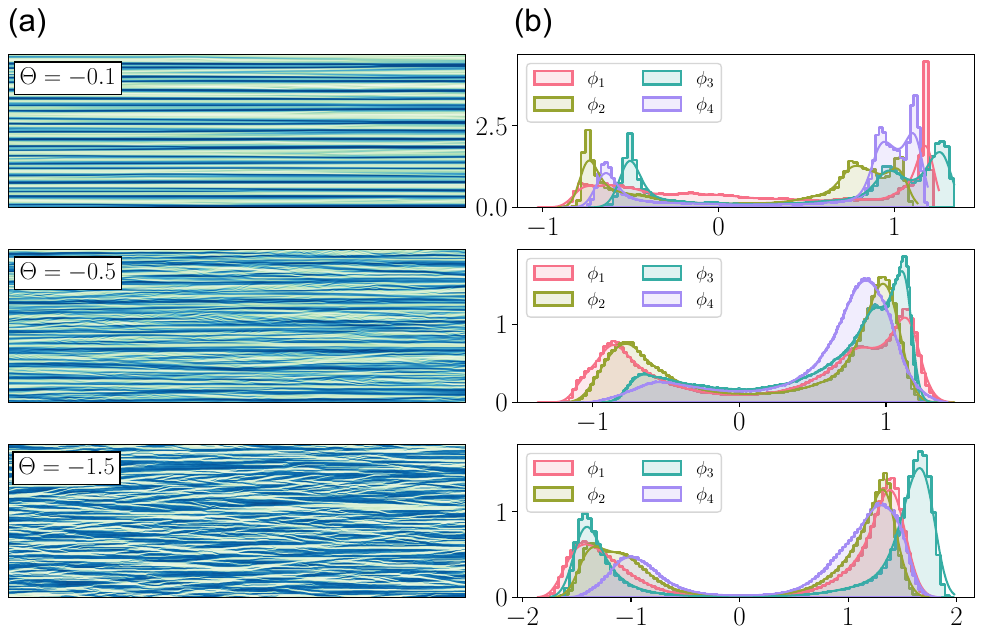}
	\caption{ \textbf{Steady state dynamics for mixture (composition drawn from a bimodal distribution):}  The left panels illustrate the dynamics in the steady state for randomly sampled realizations of $\mathbb{M}$ and $\mathbb{C}$ for a mixture with $N=100$ components. The average density of each species is chosen from a discrete bimodal distribution. i.e. it can attain two values either $d_1$ or $d_2$. The effective temperature $\Theta$ is indicated in the figure. For $\Theta$ very close to $\Theta_{\rm{sp}}$, the system undergoes bulk phase separation. As seen in the histograms of the densities of a few species selected at random, shows that species $2-4$ exhibit more than two peaks. This means that the system has compartmentalised into more than two compartments, a feature common in passive systems. At larger $\Theta$, there are two asymmetric peaks that get smoother with decreasing $\Theta$. }
	\label{fig:SimulationsBM}
\end{figure*}

\section{Imaginary eigenvalues and steady state patterns} \label{sec:StatisticsDynamics}
\noindent
In this section, we further characterize $\lambda_0$ for a finite number of components $N$ and relate the trends that we observe in the statistics of $\lambda_0$ to patterns in numerical solutions of the full nonlinear model in Eq.~\eqref{eq:multiNRCH}. $\lambda_0$ is real only when $N \to \infty$, a limit that represents a theoretical abstraction. In reality, the statistics of $\lambda_0$ exhibit strong finite-size effects that are relevant to real experimental systems and to the dynamics observed in numerical simulations. The most important aspect of this is that $\lambda_0$ is imaginary with a significant probability—an important fact because the novel features of the multi-species NRCH model can be attributed to the emergence of complex eigenvalues in the linearised dynamics of a phase-separating mixture~\cite{saha2020}.

In Fig.~\ref{fig:Stats}(a), the mean value of $\lambda_0$ (both the real and imaginary parts), averaged over 300 realisations of the matrix $\mathbb{M}$, is shown for the exponential, the discrete bimodal, and the identical distributions. In panel (b) of the same figure, the probability of $\lambda_0$ being complex is plotted for the same ensemble. A few comments are in order. First, the variance of $\lambda_0$, indicated by the error bars, shows that finite-size effects are larger for the first two cases in comparison with the identical distribution. Secondly, in the absence of compositional disorder, $\mathbb{D}$ has almost all imaginary eigenvalues for $\alpha \to 1$. The fraction of complex $\lambda_0$ approaches unity as $N$ approaches infinity, a relevant regime relevant to bio-condensates, as seen in Fig.~\ref{fig:Stats}(b). In the other two cases, the varying diagonal entries of $\mathbb{D}$ create the possibility that $\lambda_0$ is real, meaning that the probability that $\lambda_0$ is imaginary is capped by a limiting value, as seen in Fig.~\ref{fig:Stats}(b). For the exponential distribution, we expect that as $d \to 0$, or as the range of allowed values of $\bar{\phi}$ shrinks, the curve will bend upward and resemble the case with no compositional disorder.

\begin{figure*}
	\centering
	\includegraphics[width= 0.99\linewidth]{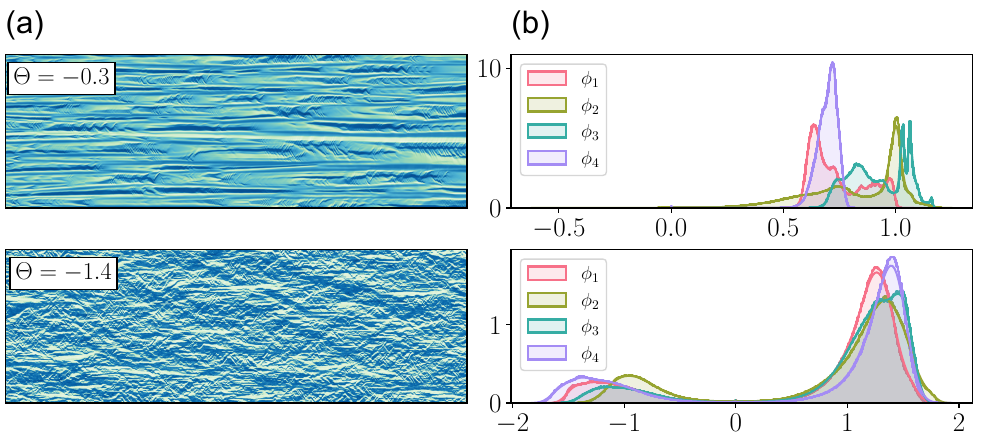}
	\caption{\textbf{Steady state dynamics for mixture (composition drawn from an exponential distribution):} The left panels illustrate the dynamics in the steady state with the diagonal disorder from an exponential distribution. Each kymograph shows the evolution of the scalar field for one species among $N$ of them. At $\Theta$ close to $\Theta_{\rm sp}$, the dynamics are chaotic, resembling chaotic condensates. As $\Theta$ is decreased, the chaotic nature of the condensates increases. The dynamics is reflected in the histograms in panel (b) where peaks in the distribution of the values assumed by the fields changes to smooth curves with two well-defined peaks.   }
	\label{fig:SimulationsExp}
\end{figure*}

We now show a few examples of the dynamics in the steady state for values of $\Theta$ chosen to be below $\Theta_{\rm sp}$. In Figs.~\ref{fig:SimulationsBM}, Figs.~\ref{fig:SimulationsExp} illustrate the dynamics in the steady state for the bimodal distribution, and the exponential distribution as $\Theta$ is tuned below $\Theta_{\rm sp}$. The same interaction matrix $\mathbb{M}$ has been chosen for the examples in panel (a) of both Figs.~\ref{fig:SimulationsBM} and \ref{fig:SimulationsExp}, the other relevant parameters are mentioned in the captions.

The effective temperature $\Theta$ is lowered (see values indicated in the panels) to showcase the dependence of pattern on the difference $\Theta_{\rm sp} - \Theta$. The kymographs representing the space-time dynamics for a randomly selected species in panel (a) of Fig.~\ref{fig:SimulationsBM} shows that just below $\Theta_{rm sp}$ condensates appear. The term condensates refers to spatial domains of enhanced or diminished density that are observed that are long lived (being stationary in the steady state). For $\Theta = -0.1$, the system seems to have arrested, a consequence of the dimensionality, and deterministic evolution of the system. On lowering the value of $\Theta$, the condensates become dynamic, notice that they shift and fluctuate in time. On increasing $\Theta$ further, we generically find spatiotemporal chaos in the steady state.

To ensure that the system has reached steady state we calculate the distribution of the values assumed by the fields in the steady state at different points in the lattice and average the data temporally as well to obtain the histograms in panel (b) of both Figs~\ref{fig:SimulationsBM} and \ref{fig:SimulationsExp}. The formation of two sharp peaks would signal binary phase separation, while smooth peaks would correlate with spatiotemporally chaotic dynamics. Notice the correlation between panels (a) and (b), stationary condensates are associated with multiple well-defined peaks in the histogram, while chaotic ones are associated with smooth and wide peaks. We note that the solid lines in the histogram plots are not a parametric fit assuming any functional form but a result of data-driven smoothing to help in distinguishing the peaks. Interestingly, we find that the condensates are likely to represent multiphase coexistence, as seen in the three or more prominent peaks are observed for $\Theta_{\rm sp}$ just below $\Theta_{\rm sp}$.

Finally, A comparison between the dynamics in Figs.~\ref{fig:SimulationsBM} and Fig.~\ref{fig:SimulationsExp} shows that similar trends appear in both cases although the transition to a chaotic condensates occurs at a larger value of $\Theta$ for the exponential distribution. Although we have reported a few examples here, further simulations have shown that the trends are quite generic when other realisations of the interaction matrix $\mathbb M$ are considered.

\section{Conclusions}\label{sec:Conclusions}
\noindent
To conclude, we have explored the role of compositional disorder in determining the onset of the spinodal instability and pattern formation in a random nonreciprocally interacting mixture. The spinodal threshold (the temperature at which the system transitions from a homogeneous to a patterned state) in a system with varying mean density of each component decreases quadratically with the strength of the nonreciprocal parameter. The stability is thus governed by the interplay of two parameters. The first is the coefficient multiplying the square of the nonreciprocal activity, and the second is the transition point at zero nonreciprocity. Both parameters are determined solely by the distribution from which the disorder is drawn. Using analytical arguments based on random matrix theory, we have motivated this functional dependence and provided arguments for the relative stability of the active system compared to its passive counterpart.

These predictions are validated through extensive numerical simulations probing whether the onset of pattern formation follows the theoretical expectations. A second set of simulations was carried out to explore the emergent features of the steady-state dynamics. We find that condensates (static domains of enhanced density) appear close to the threshold of the spinodal instability and exhibit signatures of multi-phase coexistence, meaning that the system separates into more than two dominant phases, as evidenced by multiple peaks in the distribution of density values in the steady state. On lowering the effective temperature the condensates become chaotic, a feature common to both types of compositional explored in the paper.

In passive multicomponent systems, tuning the mean density of each species generally leads to complex phase behaviour, with changes in composition providing access to new critical points~\cite{Sollich2001}. It is therefore striking that relatively simple statements can be made about the stability of the nonreciprocal system (arguably a much more complex system) when one considers an ensemble of systems with interaction strengths drawn from random distributions. The generality of the results in this realistic setting, with many components each having a variable average density, suggests that tuning the level of nonreciprocity can be an effective way to control the nature of condensates in biological systems~\cite{matsuzawa2026metabolites}. However, it is reasonable to expect that controlling the volume fractions of chemical species in a mixture, and thereby regulating phase separation, may provide a more accessible experimental pathway~\cite{Adame-Arana2020}.


Our work represents a first step toward understanding the composition dependence of multicomponent active mixtures, addressing the important aspect that different species are mixed in asymmetric proportions in realistic experimental setups and real biomolecular condensates. The detailed phase behaviour via a construction of binodals, and an implementation of Maxwell construction will be left for future work~\cite{saha2024phasecoexistencenonreciprocalcahnhilliard}. A detailed exploration of the resulting dynamical phases in two and three dimensions, where novel interfacial effects are expected upon changing the concentrations of the components~\cite{saha2024phasecoexistencenonreciprocalcahnhilliard}, will be presented in future studies.
\\\\

\begin{acknowledgements}
LP and SS acknowledge useful discussions and support from Ramin Golestanian and Navdeep Rana.  
\end{acknowledgements}

\newpage 
\section{Appendix I} \label{App:Boundary}
In this Appendix, we will describe the methods used to obtain Eq.~\eqref{eq:slope} of the main text. The resolvent for the dynamical matrix $\mathbb{D}$ is defined as
\begin{equation}
G(z)=\int \mathrm{~d} z^{\prime} \frac{\rho\left(z^{\prime}\right)}{z^{\prime}-z} 
\end{equation}
where $\rho(z)$ is the spectral density of $\mathbb{D}$ and the integral is carried out in the complex plane. As the resolvent is singular whenever $z$ is equal to one of the eigenvalues, we have to regularize the resolvent by considering the quaternionic resolvent $G(q)$, as
a function of the quaternionic argument $q$ \cite{Rogers_2010, barabas2017self}.
\begin{equation}
G(q)=\int \mathrm{~d} z^{\prime} \rho\left(z^{\prime}\right)\left(z^{\prime}-q\right)^{-1} 
\end{equation}
which is well-defined as long as $q= z + uj$ with $u \neq 0$. Using the generalization of Pastur’s relation \cite{pastur1972spectrum} for the asymmetrical matrices which was studied in \cite{Rogers_2010} and the properties of the quaternions, we can rewrite the resolvent as
\begin{equation}\label{eq:ComplexPastur}
g+\beta \mathrm{j}=\int \mathrm{~d} \bar{\phi} P(\bar{\phi}) \frac{(\bar{\phi}-{z^*}-(1-2\alpha^2) g^* ) +\beta \mathrm{j}}{|\bar{\phi}-z-(1-2\alpha^2) g|^2+|\beta|^2} ,
\end{equation}
where both sides of this equation are quaternionic. Writing the complex number $g$ as $g = \text{Re}(g)+ \text{Im}(g)$ distinguishing the real and the imaginary parts, and $z = x + yi$, we can distinguish the real and imaginary parts of Eq.~\eqref{eq:ComplexPastur} to obtain the coupled equations for $z$ and $\mbox{Re}(g)$, see Eqs.~\eqref{eq:Boundary1} and \eqref{eq:Boundary2}, in the main text. In arriving at Eqs.~\eqref{eq:Boundary1} and \eqref{eq:Boundary2} we have further assumed that the diagonal entries of $\mathbb{C}$, and thus the values $\bar{\phi}$ assume are real. 

\subsection{Identical distribution}\label{App:Identical}
\noindent
For $P = \delta(\bar{\phi} - d)$, Eqs.~\eqref{eq:Boundary1} and \eqref{eq:Boundary2} reduce to 
\beq
&& \frac{\sigma^2/4}{(\tilde{x}-d)^2 + \tilde{y}^2} = 1, \nonumber \\ 
&& \mbox{Re}(g) = \frac{d-\tilde{x}}{(\tilde{x}-d)^2 + \tilde{y}^2}. 
\eeq 
which yields $\tilde{x} = d-\sqrt{\sigma^2/4 - \tilde{y}^2}$, such that for $\tilde y = 0$, is $\tilde{x} = d - \sigma/{2}$. $\mbox{Re}(g) = 2/\sigma$. Using the definition of $\tilde{x}$. we have
\beq 
\lambda_0 = d - \sigma (1-\alpha^2).
\eeq

\subsection{Discrete Bimodal distribution}\label{App:Bimodal}
\noindent
As the second example, we sample the diagonal entries of $\mathbb{C}$ from the bimodal distribution.
$$
P(\bar{\phi})=p \delta\left(\bar{\phi}-d_1\right)+(1-p) \delta\left(\bar{\phi}-d_2\right). 
$$
We assume $d_1<d_2$ and $d_1>0$. Eqs.~\ref{eq:Boundary1} and \ref{eq:Boundary2} assume simple forms for this choice, such that $\lambda_0$ can be found analytically setting $y=0$, and using the re-defined variable $\tilde{x}$ we get
\begin{equation}
\begin{gathered}
1=\frac{p}{\left(\tilde{x}-d_1\right)^2}+\frac{(1-p)}{\left(\tilde{x}-d_2\right)^2} \\
\operatorname{Re}\left(g\right)=-\frac{p}{\left(\tilde{x}-d_1\right)}-\frac{(1-p)}{\left(\tilde{x}-d_2\right)}
\end{gathered}
\end{equation}
The smallest eigenvalue is then obtained as $\lambda_0 = \tilde{x}-d_2-(1-2\alpha^2) \sigma^2\text{Re}(g)$. However, it is worth mentioning that the boundary of support plotted in the \ref{fig:Boundary} is for $\alpha= 1/\sqrt{2}$ for which the boundary follows a simpler equation
\begin{equation}
    f(\lambda) = \frac{p}{|\lambda-d_1|^2}+\frac{(1-p)}{|\lambda-d_2|^2} = 1/\sigma^2
\end{equation}
where $\lambda=\mbox{Re}(\lambda)+i \mbox{Im}(\lambda)$.

\section{Appendix II: numerical schemes and parameter choices} \label{App:Numerics}
In this section, we will describe the methods used to solve the $N$ nonlinearly coupled PDEs to verify the results of the linear stability analysis and to garner some idea of the rich pattern forming ability of these systems of equations. The simulations shown in Fig. \ref{fig:Schematic} and used to obtain the results in Fig. \ref{fig:Simulations1} have been performed using a pseudo-spectral method in a one-dimensional system on a domain of length $L$ discretized with $nx$ grid points to obtain a stable solution of the nonlinear partial differential equations for the  N-species equations of motion. For time discretizing, we use the second-order Exponential Time Differencing (ETD2) scheme \cite{cox2002exponential} with a fixed time step $\Delta t$, a small enough time step to ensure reaching a steady state.

For the results in the Fig.~\ref{fig:Simulations1}, the simulation was carried out for $N=100$ components. The system size is chosen to be $L=4096$ and $nx=4096$ in the simulations. The time step for simulations reported in this figure is $\Delta t=2.5\times10^{-3}$.

The results of the Figs.~\ref{fig:SimulationsBM} and \ref{fig:SimulationsExp}, are from a simulation of $N=100$ components with the system size $L=512$ and $nx=512$ in the simulations. The time step for simulations reported in this figure is $\Delta t=5\times10^{-3}$. For both results in Fig. (2) and (6) the surface tension parameter $K = 0.1$. In all simulations for the discrete bimodal distribution in we chose $d_1=0.2$ and $d_2=1$. Accordingly, for the exponential distribution that was used in the simulation, we chose $d=0.6$. In Fig. (6), for the discrete bimodal distribution, from the linear stability analysis, we can find $\Theta_{sp} =0.06 $ and for the exponential distribution is $\Theta_{sp} = -0.19$.

\bibliography{biblio_Comp}

\begin{thebibliography}{81}%
\makeatletter
\providecommand \@ifxundefined [1]{%
 \@ifx{#1\undefined}
}%
\providecommand \@ifnum [1]{%
 \ifnum #1\expandafter \@firstoftwo
 \else \expandafter \@secondoftwo
 \fi
}%
\providecommand \@ifx [1]{%
 \ifx #1\expandafter \@firstoftwo
 \else \expandafter \@secondoftwo
 \fi
}%
\providecommand \natexlab [1]{#1}%
\providecommand \enquote  [1]{``#1''}%
\providecommand \bibnamefont  [1]{#1}%
\providecommand \bibfnamefont [1]{#1}%
\providecommand \citenamefont [1]{#1}%
\providecommand \href@noop [0]{\@secondoftwo}%
\providecommand \href [0]{\begingroup \@sanitize@url \@href}%
\providecommand \@href[1]{\@@startlink{#1}\@@href}%
\providecommand \@@href[1]{\endgroup#1\@@endlink}%
\providecommand \@sanitize@url [0]{\catcode `\\12\catcode `\$12\catcode
  `\&12\catcode `\#12\catcode `\^12\catcode `\_12\catcode `\%12\relax}%
\providecommand \@@startlink[1]{}%
\providecommand \@@endlink[0]{}%
\providecommand \url  [0]{\begingroup\@sanitize@url \@url }%
\providecommand \@url [1]{\endgroup\@href {#1}{\urlprefix }}%
\providecommand \urlprefix  [0]{URL }%
\providecommand \Eprint [0]{\href }%
\providecommand \doibase [0]{https://doi.org/}%
\providecommand \selectlanguage [0]{\@gobble}%
\providecommand \bibinfo  [0]{\@secondoftwo}%
\providecommand \bibfield  [0]{\@secondoftwo}%
\providecommand \translation [1]{[#1]}%
\providecommand \BibitemOpen [0]{}%
\providecommand \bibitemStop [0]{}%
\providecommand \bibitemNoStop [0]{.\EOS\space}%
\providecommand \EOS [0]{\spacefactor3000\relax}%
\providecommand \BibitemShut  [1]{\csname bibitem#1\endcsname}%
\let\auto@bib@innerbib\@empty
\bibitem [{\citenamefont {Brangwynne}\ \emph {et~al.}(2009)\citenamefont
  {Brangwynne}, \citenamefont {Eckmann}, \citenamefont {Courson}, \citenamefont
  {Rybarska}, \citenamefont {Hoege}, \citenamefont {Gharakhani}, \citenamefont
  {Jülicher},\ and\ \citenamefont {Hyman}}]{CliffBrangwynne_2009}%
  \BibitemOpen
  \bibfield  {author} {\bibinfo {author} {\bibfnamefont {C.~P.}\ \bibnamefont
  {Brangwynne}}, \bibinfo {author} {\bibfnamefont {C.~R.}\ \bibnamefont
  {Eckmann}}, \bibinfo {author} {\bibfnamefont {D.~S.}\ \bibnamefont
  {Courson}}, \bibinfo {author} {\bibfnamefont {A.}~\bibnamefont {Rybarska}},
  \bibinfo {author} {\bibfnamefont {C.}~\bibnamefont {Hoege}}, \bibinfo
  {author} {\bibfnamefont {J.}~\bibnamefont {Gharakhani}}, \bibinfo {author}
  {\bibfnamefont {F.}~\bibnamefont {Jülicher}},\ and\ \bibinfo {author}
  {\bibfnamefont {A.~A.}\ \bibnamefont {Hyman}},\ }\bibfield  {title} {\bibinfo
  {title} {Germline p granules are liquid droplets that localize by controlled
  dissolution/condensation},\ }\href {https://doi.org/10.1126/science.1172046}
  {\bibfield  {journal} {\bibinfo  {journal} {Science}\ }\textbf {\bibinfo
  {volume} {324}},\ \bibinfo {pages} {1729} (\bibinfo {year} {2009})},\ \Eprint
  {https://arxiv.org/abs/https://www.science.org/doi/pdf/10.1126/science.1172046}
  {https://www.science.org/doi/pdf/10.1126/science.1172046} \BibitemShut
  {NoStop}%
\bibitem [{\citenamefont {Alberti}\ and\ \citenamefont
  {Hyman}(2021)}]{Alberti2021}%
  \BibitemOpen
  \bibfield  {author} {\bibinfo {author} {\bibfnamefont {S.}~\bibnamefont
  {Alberti}}\ and\ \bibinfo {author} {\bibfnamefont {A.~A.}\ \bibnamefont
  {Hyman}},\ }\bibfield  {title} {\bibinfo {title} {Biomolecular condensates at
  the nexus of cellular stress, protein aggregation disease and ageing},\
  }\href {https://doi.org/10.1038/s41580-020-00326-6} {\bibfield  {journal}
  {\bibinfo  {journal} {Nature Reviews Molecular Cell Biology}\ }\textbf
  {\bibinfo {volume} {22}},\ \bibinfo {pages} {196} (\bibinfo {year}
  {2021})}\BibitemShut {NoStop}%
\bibitem [{\citenamefont {Hyman}\ \emph {et~al.}(2014)\citenamefont {Hyman},
  \citenamefont {Weber},\ and\ \citenamefont {Jülicher}}]{HymanReview}%
  \BibitemOpen
  \bibfield  {author} {\bibinfo {author} {\bibfnamefont {A.~A.}\ \bibnamefont
  {Hyman}}, \bibinfo {author} {\bibfnamefont {C.~A.}\ \bibnamefont {Weber}},\
  and\ \bibinfo {author} {\bibfnamefont {F.}~\bibnamefont {Jülicher}},\
  }\bibfield  {title} {\bibinfo {title} {Liquid-liquid phase separation in
  biology},\ }\href
  {https://doi.org/https://doi.org/10.1146/annurev-cellbio-100913-013325}
  {\bibfield  {journal} {\bibinfo  {journal} {Annual Review of Cell and
  Developmental Biology}\ }\textbf {\bibinfo {volume} {30}},\ \bibinfo {pages}
  {39} (\bibinfo {year} {2014})}\BibitemShut {NoStop}%
\bibitem [{\citenamefont {Shin}\ and\ \citenamefont
  {Brangwynne}(2017)}]{shin2017liquid}%
  \BibitemOpen
  \bibfield  {author} {\bibinfo {author} {\bibfnamefont {Y.}~\bibnamefont
  {Shin}}\ and\ \bibinfo {author} {\bibfnamefont {C.~P.}\ \bibnamefont
  {Brangwynne}},\ }\bibfield  {title} {\bibinfo {title} {Liquid phase
  condensation in cell physiology and disease},\ }\href@noop {} {\bibfield
  {journal} {\bibinfo  {journal} {Science}\ }\textbf {\bibinfo {volume}
  {357}},\ \bibinfo {pages} {eaaf4382} (\bibinfo {year} {2017})}\BibitemShut
  {NoStop}%
\bibitem [{\citenamefont {Cohan}\ and\ \citenamefont
  {Pappu}(2020)}]{cohan2020making}%
  \BibitemOpen
  \bibfield  {author} {\bibinfo {author} {\bibfnamefont {M.~C.}\ \bibnamefont
  {Cohan}}\ and\ \bibinfo {author} {\bibfnamefont {R.~V.}\ \bibnamefont
  {Pappu}},\ }\bibfield  {title} {\bibinfo {title} {Making the case for
  disordered proteins and biomolecular condensates in bacteria},\ }\href@noop
  {} {\bibfield  {journal} {\bibinfo  {journal} {Trends in Biochemical
  Sciences}\ }\textbf {\bibinfo {volume} {45}},\ \bibinfo {pages} {668}
  (\bibinfo {year} {2020})}\BibitemShut {NoStop}%
\bibitem [{\citenamefont {Brangwynne}\ \emph {et~al.}(2015)\citenamefont
  {Brangwynne}, \citenamefont {Tompa},\ and\ \citenamefont
  {Pappu}}]{brangwynne2015polymer}%
  \BibitemOpen
  \bibfield  {author} {\bibinfo {author} {\bibfnamefont {C.~P.}\ \bibnamefont
  {Brangwynne}}, \bibinfo {author} {\bibfnamefont {P.}~\bibnamefont {Tompa}},\
  and\ \bibinfo {author} {\bibfnamefont {R.~V.}\ \bibnamefont {Pappu}},\
  }\bibfield  {title} {\bibinfo {title} {Polymer physics of intracellular phase
  transitions},\ }\href@noop {} {\bibfield  {journal} {\bibinfo  {journal}
  {Nature Physics}\ }\textbf {\bibinfo {volume} {11}},\ \bibinfo {pages} {899}
  (\bibinfo {year} {2015})}\BibitemShut {NoStop}%
\bibitem [{\citenamefont {Iarovaia}\ \emph {et~al.}(2019)\citenamefont
  {Iarovaia}, \citenamefont {Minina}, \citenamefont {Sheval}, \citenamefont
  {Onichtchouk}, \citenamefont {Dokudovskaya}, \citenamefont {Razin},\ and\
  \citenamefont {Vassetzky}}]{Iarovaia2019}%
  \BibitemOpen
  \bibfield  {author} {\bibinfo {author} {\bibfnamefont {O.~V.}\ \bibnamefont
  {Iarovaia}}, \bibinfo {author} {\bibfnamefont {E.~P.}\ \bibnamefont
  {Minina}}, \bibinfo {author} {\bibfnamefont {E.~V.}\ \bibnamefont {Sheval}},
  \bibinfo {author} {\bibfnamefont {D.}~\bibnamefont {Onichtchouk}}, \bibinfo
  {author} {\bibfnamefont {S.}~\bibnamefont {Dokudovskaya}}, \bibinfo {author}
  {\bibfnamefont {S.~V.}\ \bibnamefont {Razin}},\ and\ \bibinfo {author}
  {\bibfnamefont {Y.~S.}\ \bibnamefont {Vassetzky}},\ }\bibfield  {title}
  {\bibinfo {title} {Nucleolus: A central hub for nuclear functions},\ }\href
  {https://www.sciencedirect.com/science/article/pii/S0962892419300789}
  {\bibfield  {journal} {\bibinfo  {journal} {Trends in Cell Biology}\ }\textbf
  {\bibinfo {volume} {29}},\ \bibinfo {pages} {647} (\bibinfo {year}
  {2019})}\BibitemShut {NoStop}%
\bibitem [{\citenamefont {Brangwynne}\ \emph {et~al.}(2011)\citenamefont
  {Brangwynne}, \citenamefont {Mitchison},\ and\ \citenamefont
  {Hyman}}]{brangwynne2011active}%
  \BibitemOpen
  \bibfield  {author} {\bibinfo {author} {\bibfnamefont {C.~P.}\ \bibnamefont
  {Brangwynne}}, \bibinfo {author} {\bibfnamefont {T.~J.}\ \bibnamefont
  {Mitchison}},\ and\ \bibinfo {author} {\bibfnamefont {A.~A.}\ \bibnamefont
  {Hyman}},\ }\bibfield  {title} {\bibinfo {title} {Active liquid-like behavior
  of nucleoli determines their size and shape in xenopus laevis oocytes},\
  }\href@noop {} {\bibfield  {journal} {\bibinfo  {journal} {Proceedings of the
  National Academy of Sciences}\ }\textbf {\bibinfo {volume} {108}},\ \bibinfo
  {pages} {4334} (\bibinfo {year} {2011})}\BibitemShut {NoStop}%
\bibitem [{\citenamefont {Lafontaine}\ \emph {et~al.}(2021)\citenamefont
  {Lafontaine}, \citenamefont {Riback}, \citenamefont {Bascetin},\ and\
  \citenamefont {Brangwynne}}]{lafontaine2021nucleolus}%
  \BibitemOpen
  \bibfield  {author} {\bibinfo {author} {\bibfnamefont {D.~L.}\ \bibnamefont
  {Lafontaine}}, \bibinfo {author} {\bibfnamefont {J.~A.}\ \bibnamefont
  {Riback}}, \bibinfo {author} {\bibfnamefont {R.}~\bibnamefont {Bascetin}},\
  and\ \bibinfo {author} {\bibfnamefont {C.~P.}\ \bibnamefont {Brangwynne}},\
  }\bibfield  {title} {\bibinfo {title} {The nucleolus as a multiphase liquid
  condensate},\ }\href@noop {} {\bibfield  {journal} {\bibinfo  {journal}
  {Nature reviews Molecular cell biology}\ }\textbf {\bibinfo {volume} {22}},\
  \bibinfo {pages} {165} (\bibinfo {year} {2021})}\BibitemShut {NoStop}%
\bibitem [{\citenamefont {Protter}\ and\ \citenamefont
  {Parker}(2016)}]{Protter2016}%
  \BibitemOpen
  \bibfield  {author} {\bibinfo {author} {\bibfnamefont {D.~S.}\ \bibnamefont
  {Protter}}\ and\ \bibinfo {author} {\bibfnamefont {R.}~\bibnamefont
  {Parker}},\ }\bibfield  {title} {\bibinfo {title} {Principles and properties
  of stress granules},\ }\href
  {https://www.sciencedirect.com/science/article/pii/S0962892416300472}
  {\bibfield  {journal} {\bibinfo  {journal} {Trends in Cell Biology}\ }\textbf
  {\bibinfo {volume} {26}},\ \bibinfo {pages} {668} (\bibinfo {year}
  {2016})}\BibitemShut {NoStop}%
\bibitem [{\citenamefont {Franzmann}\ \emph {et~al.}(2018)\citenamefont
  {Franzmann}, \citenamefont {Jahnel}, \citenamefont {Pozniakovsky},
  \citenamefont {Mahamid}, \citenamefont {Holehouse}, \citenamefont
  {N{\"u}ske}, \citenamefont {Richter}, \citenamefont {Baumeister},
  \citenamefont {Grill}, \citenamefont {Pappu}, \citenamefont {Hyman},\ and\
  \citenamefont {Alberti}}]{Franzmann2018}%
  \BibitemOpen
  \bibfield  {author} {\bibinfo {author} {\bibfnamefont {T.~M.}\ \bibnamefont
  {Franzmann}}, \bibinfo {author} {\bibfnamefont {M.}~\bibnamefont {Jahnel}},
  \bibinfo {author} {\bibfnamefont {A.}~\bibnamefont {Pozniakovsky}}, \bibinfo
  {author} {\bibfnamefont {J.}~\bibnamefont {Mahamid}}, \bibinfo {author}
  {\bibfnamefont {A.~S.}\ \bibnamefont {Holehouse}}, \bibinfo {author}
  {\bibfnamefont {E.}~\bibnamefont {N{\"u}ske}}, \bibinfo {author}
  {\bibfnamefont {D.}~\bibnamefont {Richter}}, \bibinfo {author} {\bibfnamefont
  {W.}~\bibnamefont {Baumeister}}, \bibinfo {author} {\bibfnamefont {S.~W.}\
  \bibnamefont {Grill}}, \bibinfo {author} {\bibfnamefont {R.~V.}\ \bibnamefont
  {Pappu}}, \bibinfo {author} {\bibfnamefont {A.~A.}\ \bibnamefont {Hyman}},\
  and\ \bibinfo {author} {\bibfnamefont {S.}~\bibnamefont {Alberti}},\
  }\bibfield  {title} {\bibinfo {title} {Phase separation of a yeast prion
  protein promotes cellular fitness},\ }\href
  {https://doi.org/10.1126/science.aao5654} {\bibfield  {journal} {\bibinfo
  {journal} {Science}\ }\textbf {\bibinfo {volume} {359}},\ \bibinfo {pages}
  {eaao5654} (\bibinfo {year} {2018})}\BibitemShut {NoStop}%
\bibitem [{\citenamefont {Wilson}(1899)}]{EdmundWilson_1899}%
  \BibitemOpen
  \bibfield  {author} {\bibinfo {author} {\bibfnamefont {E.~B.}\ \bibnamefont
  {Wilson}},\ }\bibfield  {title} {\bibinfo {title} {The structure of
  protoplasm},\ }\href {https://doi.org/10.1126/science.10.237.33} {\bibfield
  {journal} {\bibinfo  {journal} {Science}\ }\textbf {\bibinfo {volume} {10}},\
  \bibinfo {pages} {33} (\bibinfo {year} {1899})},\ \Eprint
  {https://arxiv.org/abs/https://www.science.org/doi/pdf/10.1126/science.10.237.33}
  {https://www.science.org/doi/pdf/10.1126/science.10.237.33} \BibitemShut
  {NoStop}%
\bibitem [{\citenamefont {Wang}\ \emph {et~al.}(2021)\citenamefont {Wang},
  \citenamefont {Zhang}, \citenamefont {Dai}, \citenamefont {Qin},
  \citenamefont {Lu}, \citenamefont {Zhang},\ and\ \citenamefont
  {Zhou}}]{Wang2021}%
  \BibitemOpen
  \bibfield  {author} {\bibinfo {author} {\bibfnamefont {B.}~\bibnamefont
  {Wang}}, \bibinfo {author} {\bibfnamefont {L.}~\bibnamefont {Zhang}},
  \bibinfo {author} {\bibfnamefont {T.}~\bibnamefont {Dai}}, \bibinfo {author}
  {\bibfnamefont {Z.}~\bibnamefont {Qin}}, \bibinfo {author} {\bibfnamefont
  {H.}~\bibnamefont {Lu}}, \bibinfo {author} {\bibfnamefont {L.}~\bibnamefont
  {Zhang}},\ and\ \bibinfo {author} {\bibfnamefont {F.}~\bibnamefont {Zhou}},\
  }\bibfield  {title} {\bibinfo {title} {Liquid--liquid phase separation in
  human health and diseases},\ }\href
  {https://doi.org/10.1038/s41392-021-00678-1} {\bibfield  {journal} {\bibinfo
  {journal} {Signal Transduction and Targeted Therapy}\ }\textbf {\bibinfo
  {volume} {6}},\ \bibinfo {pages} {290} (\bibinfo {year} {2021})}\BibitemShut
  {NoStop}%
\bibitem [{\citenamefont {Oparin}(2003)}]{oparin2003origin}%
  \BibitemOpen
  \bibfield  {author} {\bibinfo {author} {\bibfnamefont {A.}~\bibnamefont
  {Oparin}},\ }\href {https://books.google.de/books?id=wQE_yQEACAAJ} {\emph
  {\bibinfo {title} {The Origin of Life}}},\ Dover phoenix editions\ (\bibinfo
  {publisher} {Dover Publications},\ \bibinfo {year} {2003})\BibitemShut
  {NoStop}%
\bibitem [{\citenamefont {Adame-Arana}\ \emph {et~al.}(2020)\citenamefont
  {Adame-Arana}, \citenamefont {Weber}, \citenamefont {Zaburdaev},
  \citenamefont {Prost},\ and\ \citenamefont {J{\"u}licher}}]{Adame-Arana2020}%
  \BibitemOpen
  \bibfield  {author} {\bibinfo {author} {\bibfnamefont {O.}~\bibnamefont
  {Adame-Arana}}, \bibinfo {author} {\bibfnamefont {C.~A.}\ \bibnamefont
  {Weber}}, \bibinfo {author} {\bibfnamefont {V.}~\bibnamefont {Zaburdaev}},
  \bibinfo {author} {\bibfnamefont {J.}~\bibnamefont {Prost}},\ and\ \bibinfo
  {author} {\bibfnamefont {F.}~\bibnamefont {J{\"u}licher}},\ }\bibfield
  {title} {\bibinfo {title} {Liquid phase separation controlled by ph},\ }\href
  {https://doi.org/10.1016/j.bpj.2020.07.044} {\bibfield  {journal} {\bibinfo
  {journal} {Biophysical Journal}\ }\textbf {\bibinfo {volume} {119}},\
  \bibinfo {pages} {1590} (\bibinfo {year} {2020})}\BibitemShut {NoStop}%
\bibitem [{\citenamefont {Patel}\ \emph {et~al.}(2017)\citenamefont {Patel},
  \citenamefont {Malinovska}, \citenamefont {Saha}, \citenamefont {Wang},
  \citenamefont {Alberti}, \citenamefont {Krishnan},\ and\ \citenamefont
  {Hyman}}]{Patel_ATP_LLPS}%
  \BibitemOpen
  \bibfield  {author} {\bibinfo {author} {\bibfnamefont {A.}~\bibnamefont
  {Patel}}, \bibinfo {author} {\bibfnamefont {L.}~\bibnamefont {Malinovska}},
  \bibinfo {author} {\bibfnamefont {S.}~\bibnamefont {Saha}}, \bibinfo {author}
  {\bibfnamefont {J.}~\bibnamefont {Wang}}, \bibinfo {author} {\bibfnamefont
  {S.}~\bibnamefont {Alberti}}, \bibinfo {author} {\bibfnamefont
  {Y.}~\bibnamefont {Krishnan}},\ and\ \bibinfo {author} {\bibfnamefont
  {A.~A.}\ \bibnamefont {Hyman}},\ }\bibfield  {title} {\bibinfo {title} {Atp
  as a biological hydrotrope},\ }\href
  {https://doi.org/10.1126/science.aaf6846} {\bibfield  {journal} {\bibinfo
  {journal} {Science}\ }\textbf {\bibinfo {volume} {356}},\ \bibinfo {pages}
  {753} (\bibinfo {year} {2017})},\ \Eprint
  {https://arxiv.org/abs/https://www.science.org/doi/pdf/10.1126/science.aaf6846}
  {https://www.science.org/doi/pdf/10.1126/science.aaf6846} \BibitemShut
  {NoStop}%
\bibitem [{\citenamefont {Weber}\ \emph {et~al.}(2017)\citenamefont {Weber},
  \citenamefont {Lee},\ and\ \citenamefont {Jülicher}}]{Weber_2017}%
  \BibitemOpen
  \bibfield  {author} {\bibinfo {author} {\bibfnamefont {C.~A.}\ \bibnamefont
  {Weber}}, \bibinfo {author} {\bibfnamefont {C.~F.}\ \bibnamefont {Lee}},\
  and\ \bibinfo {author} {\bibfnamefont {F.}~\bibnamefont {Jülicher}},\
  }\bibfield  {title} {\bibinfo {title} {Droplet ripening in concentration
  gradients},\ }\href {https://doi.org/10.1088/1367-2630/aa6b84} {\bibfield
  {journal} {\bibinfo  {journal} {New Journal of Physics}\ }\textbf {\bibinfo
  {volume} {19}},\ \bibinfo {pages} {053021} (\bibinfo {year}
  {2017})}\BibitemShut {NoStop}%
\bibitem [{\citenamefont {Tayar}\ \emph {et~al.}(2023)\citenamefont {Tayar},
  \citenamefont {Caballero}, \citenamefont {Anderberg}, \citenamefont {Saleh},
  \citenamefont {Cristina~Marchetti},\ and\ \citenamefont {Dogic}}]{Tayar2023}%
  \BibitemOpen
  \bibfield  {author} {\bibinfo {author} {\bibfnamefont {A.~M.}\ \bibnamefont
  {Tayar}}, \bibinfo {author} {\bibfnamefont {F.}~\bibnamefont {Caballero}},
  \bibinfo {author} {\bibfnamefont {T.}~\bibnamefont {Anderberg}}, \bibinfo
  {author} {\bibfnamefont {O.~A.}\ \bibnamefont {Saleh}}, \bibinfo {author}
  {\bibfnamefont {M.}~\bibnamefont {Cristina~Marchetti}},\ and\ \bibinfo
  {author} {\bibfnamefont {Z.}~\bibnamefont {Dogic}},\ }\bibfield  {title}
  {\bibinfo {title} {Controlling liquid--liquid phase behaviour with an active
  fluid},\ }\href {https://doi.org/10.1038/s41563-023-01660-8} {\bibfield
  {journal} {\bibinfo  {journal} {Nature Materials}\ }\textbf {\bibinfo
  {volume} {22}},\ \bibinfo {pages} {1401} (\bibinfo {year}
  {2023})}\BibitemShut {NoStop}%
\bibitem [{\citenamefont {Sollich}(2001{\natexlab{a}})}]{PeterSollich_2002}%
  \BibitemOpen
  \bibfield  {author} {\bibinfo {author} {\bibfnamefont {P.}~\bibnamefont
  {Sollich}},\ }\bibfield  {title} {\bibinfo {title} {Predicting phase
  equilibria in polydisperse systems},\ }\href
  {https://doi.org/10.1088/0953-8984/14/3/201} {\bibfield  {journal} {\bibinfo
  {journal} {Journal of Physics: Condensed Matter}\ }\textbf {\bibinfo {volume}
  {14}},\ \bibinfo {pages} {R79} (\bibinfo {year}
  {2001}{\natexlab{a}})}\BibitemShut {NoStop}%
\bibitem [{\citenamefont {Zimmermann}(1997)}]{zimmermann1997}%
  \BibitemOpen
  \bibfield  {author} {\bibinfo {author} {\bibfnamefont {W.}~\bibnamefont
  {Zimmermann}},\ }\bibfield  {title} {\bibinfo {title} {Stability of traveling
  waves for a conserved field},\ }\href
  {https://doi.org/10.1016/S0378-4371(96)00422-0} {\bibfield  {journal}
  {\bibinfo  {journal} {Physica A: Statistical Mechanics and its Applications}\
  }\textbf {\bibinfo {volume} {237}},\ \bibinfo {pages} {405} (\bibinfo {year}
  {1997})}\BibitemShut {NoStop}%
\bibitem [{\citenamefont {You}\ \emph {et~al.}(2020{\natexlab{a}})\citenamefont
  {You}, \citenamefont {Huang}, \citenamefont {Yu}, \citenamefont {Shen},
  \citenamefont {Sevilla}, \citenamefont {Shi}, \citenamefont {Hermjakob},
  \citenamefont {Chen},\ and\ \citenamefont {Li}}]{You2020}%
  \BibitemOpen
  \bibfield  {author} {\bibinfo {author} {\bibfnamefont {K.}~\bibnamefont
  {You}}, \bibinfo {author} {\bibfnamefont {Q.}~\bibnamefont {Huang}}, \bibinfo
  {author} {\bibfnamefont {C.}~\bibnamefont {Yu}}, \bibinfo {author}
  {\bibfnamefont {B.}~\bibnamefont {Shen}}, \bibinfo {author} {\bibfnamefont
  {C.}~\bibnamefont {Sevilla}}, \bibinfo {author} {\bibfnamefont
  {M.}~\bibnamefont {Shi}}, \bibinfo {author} {\bibfnamefont {H.}~\bibnamefont
  {Hermjakob}}, \bibinfo {author} {\bibfnamefont {Y.}~\bibnamefont {Chen}},\
  and\ \bibinfo {author} {\bibfnamefont {T.}~\bibnamefont {Li}},\ }\bibfield
  {title} {\bibinfo {title} {Phasepdb: a database of liquid--liquid phase
  separation related proteins},\ }\href {https://doi.org/10.1093/nar/gkz847}
  {\bibfield  {journal} {\bibinfo  {journal} {Nucleic Acids Research}\ }\textbf
  {\bibinfo {volume} {48}},\ \bibinfo {pages} {D354} (\bibinfo {year}
  {2020}{\natexlab{a}})}\BibitemShut {NoStop}%
\bibitem [{\citenamefont {Sollich}(2001{\natexlab{b}})}]{Sollich2001}%
  \BibitemOpen
  \bibfield  {author} {\bibinfo {author} {\bibfnamefont {P.}~\bibnamefont
  {Sollich}},\ }\bibfield  {title} {\bibinfo {title} {Predicting phase
  equilibria in polydisperse systems},\ }\href
  {https://doi.org/10.1088/0953-8984/14/3/201} {\bibfield  {journal} {\bibinfo
  {journal} {Journal of Physics: Condensed Matter}\ }\textbf {\bibinfo {volume}
  {14}},\ \bibinfo {pages} {R79} (\bibinfo {year}
  {2001}{\natexlab{b}})}\BibitemShut {NoStop}%
\bibitem [{\citenamefont {Sear}\ and\ \citenamefont
  {Cuesta}(2003)}]{Sear_2003}%
  \BibitemOpen
  \bibfield  {author} {\bibinfo {author} {\bibfnamefont {R.~P.}\ \bibnamefont
  {Sear}}\ and\ \bibinfo {author} {\bibfnamefont {J.~A.}\ \bibnamefont
  {Cuesta}},\ }\bibfield  {title} {\bibinfo {title} {Instabilities in complex
  mixtures with a large number of components},\ }\bibfield  {journal} {\bibinfo
   {journal} {Physical Review Letters}\ }\textbf {\bibinfo {volume} {91}},\
  \href {https://doi.org/10.1103/physrevlett.91.245701}
  {10.1103/physrevlett.91.245701} (\bibinfo {year} {2003})\BibitemShut
  {NoStop}%
\bibitem [{\citenamefont {Jacobs}(2023)}]{jacobs2023theory}%
  \BibitemOpen
  \bibfield  {author} {\bibinfo {author} {\bibfnamefont {W.~M.}\ \bibnamefont
  {Jacobs}},\ }\bibfield  {title} {\bibinfo {title} {Theory and simulation of
  multiphase coexistence in biomolecular mixtures},\ }\href@noop {} {\bibfield
  {journal} {\bibinfo  {journal} {Journal of Chemical Theory and Computation}\
  } (\bibinfo {year} {2023})}\BibitemShut {NoStop}%
\bibitem [{\citenamefont {Jacobs}\ and\ \citenamefont
  {Frenkel}(2017)}]{jacobs2017phase}%
  \BibitemOpen
  \bibfield  {author} {\bibinfo {author} {\bibfnamefont {W.~M.}\ \bibnamefont
  {Jacobs}}\ and\ \bibinfo {author} {\bibfnamefont {D.}~\bibnamefont
  {Frenkel}},\ }\bibfield  {title} {\bibinfo {title} {Phase transitions in
  biological systems with many components},\ }\href@noop {} {\bibfield
  {journal} {\bibinfo  {journal} {Biophysical journal}\ }\textbf {\bibinfo
  {volume} {112}},\ \bibinfo {pages} {683} (\bibinfo {year}
  {2017})}\BibitemShut {NoStop}%
\bibitem [{\citenamefont {Shrinivas}\ and\ \citenamefont
  {Brenner}(2021)}]{Shrinivas2021}%
  \BibitemOpen
  \bibfield  {author} {\bibinfo {author} {\bibfnamefont {K.}~\bibnamefont
  {Shrinivas}}\ and\ \bibinfo {author} {\bibfnamefont {M.~P.}\ \bibnamefont
  {Brenner}},\ }\bibfield  {title} {\bibinfo {title} {Phase separation in
  fluids with many interacting components},\ }\href
  {https://doi.org/10.1073/pnas.2108551118} {\bibfield  {journal} {\bibinfo
  {journal} {Proceedings of the National Academy of Sciences}\ }\textbf
  {\bibinfo {volume} {118}},\ \bibinfo {pages} {e2108551118} (\bibinfo {year}
  {2021})}\BibitemShut {NoStop}%
\bibitem [{\citenamefont {Hondele}\ \emph {et~al.}(2020)\citenamefont
  {Hondele}, \citenamefont {Heinrich}, \citenamefont {De~Los~Rios},\ and\
  \citenamefont {Weis}}]{WeisHondele2020}%
  \BibitemOpen
  \bibfield  {author} {\bibinfo {author} {\bibfnamefont {M.}~\bibnamefont
  {Hondele}}, \bibinfo {author} {\bibfnamefont {S.}~\bibnamefont {Heinrich}},
  \bibinfo {author} {\bibfnamefont {P.}~\bibnamefont {De~Los~Rios}},\ and\
  \bibinfo {author} {\bibfnamefont {K.}~\bibnamefont {Weis}},\ }\bibfield
  {title} {\bibinfo {title} {Membraneless organelles: phasing out of
  equilibrium},\ }\href {https://doi.org/10.1042/ETLS20190190} {\bibfield
  {journal} {\bibinfo  {journal} {Emerging Topics in Life Sciences}\ }\textbf
  {\bibinfo {volume} {4}},\ \bibinfo {pages} {343} (\bibinfo {year}
  {2020})}\BibitemShut {NoStop}%
\bibitem [{\citenamefont {Aierken}\ \emph {et~al.}(2026)\citenamefont
  {Aierken}, \citenamefont {Aland}, \citenamefont {Bo}, \citenamefont
  {Boeynaems}, \citenamefont {Cai}, \citenamefont {Carra}, \citenamefont
  {Case}, \citenamefont {Chan}, \citenamefont {Espinosa}, \citenamefont
  {GrandPre}, \citenamefont {Grosberg}, \citenamefont {Haugerud}, \citenamefont
  {Jacobs}, \citenamefont {Joseph}, \citenamefont {Jülicher}, \citenamefont
  {Kremer}, \citenamefont {Kusters}, \citenamefont {Laan}, \citenamefont
  {Lasker}, \citenamefont {Laxhuber}, \citenamefont {Lee}, \citenamefont {Liu},
  \citenamefont {Notani}, \citenamefont {Qiang}, \citenamefont {Robustelli},
  \citenamefont {Saiz}, \citenamefont {Saleh}, \citenamefont {Schiessel},
  \citenamefont {Schmit}, \citenamefont {Shen}, \citenamefont {Shrinivas},
  \citenamefont {Statt}, \citenamefont {Tejedor}, \citenamefont {Trcek},
  \citenamefont {Weber}, \citenamefont {Weber}, \citenamefont {Wingreen},
  \citenamefont {Zhang}, \citenamefont {Zhang}, \citenamefont {Zhou},\ and\
  \citenamefont {Zwicker}}]{aierken2026roadmapcondensatescellbiology}%
  \BibitemOpen
  \bibfield  {author} {\bibinfo {author} {\bibfnamefont {D.}~\bibnamefont
  {Aierken}}, \bibinfo {author} {\bibfnamefont {S.}~\bibnamefont {Aland}},
  \bibinfo {author} {\bibfnamefont {S.}~\bibnamefont {Bo}}, \bibinfo {author}
  {\bibfnamefont {S.}~\bibnamefont {Boeynaems}}, \bibinfo {author}
  {\bibfnamefont {D.}~\bibnamefont {Cai}}, \bibinfo {author} {\bibfnamefont
  {S.}~\bibnamefont {Carra}}, \bibinfo {author} {\bibfnamefont {L.~B.}\
  \bibnamefont {Case}}, \bibinfo {author} {\bibfnamefont {H.~S.}\ \bibnamefont
  {Chan}}, \bibinfo {author} {\bibfnamefont {J.~R.}\ \bibnamefont {Espinosa}},
  \bibinfo {author} {\bibfnamefont {T.~K.}\ \bibnamefont {GrandPre}}, \bibinfo
  {author} {\bibfnamefont {A.~Y.}\ \bibnamefont {Grosberg}}, \bibinfo {author}
  {\bibfnamefont {I.~S.}\ \bibnamefont {Haugerud}}, \bibinfo {author}
  {\bibfnamefont {W.~M.}\ \bibnamefont {Jacobs}}, \bibinfo {author}
  {\bibfnamefont {J.~A.}\ \bibnamefont {Joseph}}, \bibinfo {author}
  {\bibfnamefont {F.}~\bibnamefont {Jülicher}}, \bibinfo {author}
  {\bibfnamefont {K.}~\bibnamefont {Kremer}}, \bibinfo {author} {\bibfnamefont
  {G.}~\bibnamefont {Kusters}}, \bibinfo {author} {\bibfnamefont
  {L.}~\bibnamefont {Laan}}, \bibinfo {author} {\bibfnamefont {K.}~\bibnamefont
  {Lasker}}, \bibinfo {author} {\bibfnamefont {K.~S.}\ \bibnamefont
  {Laxhuber}}, \bibinfo {author} {\bibfnamefont {H.~O.}\ \bibnamefont {Lee}},
  \bibinfo {author} {\bibfnamefont {K.~F.}\ \bibnamefont {Liu}}, \bibinfo
  {author} {\bibfnamefont {D.}~\bibnamefont {Notani}}, \bibinfo {author}
  {\bibfnamefont {Y.}~\bibnamefont {Qiang}}, \bibinfo {author} {\bibfnamefont
  {P.}~\bibnamefont {Robustelli}}, \bibinfo {author} {\bibfnamefont
  {L.}~\bibnamefont {Saiz}}, \bibinfo {author} {\bibfnamefont {O.~A.}\
  \bibnamefont {Saleh}}, \bibinfo {author} {\bibfnamefont {H.}~\bibnamefont
  {Schiessel}}, \bibinfo {author} {\bibfnamefont {J.}~\bibnamefont {Schmit}},
  \bibinfo {author} {\bibfnamefont {M.}~\bibnamefont {Shen}}, \bibinfo {author}
  {\bibfnamefont {K.}~\bibnamefont {Shrinivas}}, \bibinfo {author}
  {\bibfnamefont {A.}~\bibnamefont {Statt}}, \bibinfo {author} {\bibfnamefont
  {A.~R.}\ \bibnamefont {Tejedor}}, \bibinfo {author} {\bibfnamefont
  {T.}~\bibnamefont {Trcek}}, \bibinfo {author} {\bibfnamefont {C.~A.}\
  \bibnamefont {Weber}}, \bibinfo {author} {\bibfnamefont {S.~C.}\ \bibnamefont
  {Weber}}, \bibinfo {author} {\bibfnamefont {N.~S.}\ \bibnamefont {Wingreen}},
  \bibinfo {author} {\bibfnamefont {H.}~\bibnamefont {Zhang}}, \bibinfo
  {author} {\bibfnamefont {Y.}~\bibnamefont {Zhang}}, \bibinfo {author}
  {\bibfnamefont {H.~X.}\ \bibnamefont {Zhou}},\ and\ \bibinfo {author}
  {\bibfnamefont {D.}~\bibnamefont {Zwicker}},\ }\href
  {https://arxiv.org/abs/2601.03677} {\bibinfo {title} {Roadmap for condensates
  in cell biology}} (\bibinfo {year} {2026}),\ \Eprint
  {https://arxiv.org/abs/2601.03677} {arXiv:2601.03677 [physics.bio-ph]}
  \BibitemShut {NoStop}%
\bibitem [{\citenamefont {Cates}\ and\ \citenamefont
  {Tailleur}(2015)}]{cates2015motility}%
  \BibitemOpen
  \bibfield  {author} {\bibinfo {author} {\bibfnamefont {M.~E.}\ \bibnamefont
  {Cates}}\ and\ \bibinfo {author} {\bibfnamefont {J.}~\bibnamefont
  {Tailleur}},\ }\bibfield  {title} {\bibinfo {title} {Motility-induced phase
  separation},\ }\href
  {https://doi.org/10.1146/annurev-conmatphys-031214-014710} {\bibfield
  {journal} {\bibinfo  {journal} {Annu. Rev. Condens. Matter Phys.}\ }\textbf
  {\bibinfo {volume} {6}},\ \bibinfo {pages} {219} (\bibinfo {year}
  {2015})}\BibitemShut {NoStop}%
\bibitem [{\citenamefont {Wittkowski}\ \emph {et~al.}(2014)\citenamefont
  {Wittkowski}, \citenamefont {Tiribocchi}, \citenamefont {Stenhammar},
  \citenamefont {Allen}, \citenamefont {Marenduzzo},\ and\ \citenamefont
  {Cates}}]{Wittkowski2014}%
  \BibitemOpen
  \bibfield  {author} {\bibinfo {author} {\bibfnamefont {R.}~\bibnamefont
  {Wittkowski}}, \bibinfo {author} {\bibfnamefont {A.}~\bibnamefont
  {Tiribocchi}}, \bibinfo {author} {\bibfnamefont {J.}~\bibnamefont
  {Stenhammar}}, \bibinfo {author} {\bibfnamefont {R.~J.}\ \bibnamefont
  {Allen}}, \bibinfo {author} {\bibfnamefont {D.}~\bibnamefont {Marenduzzo}},\
  and\ \bibinfo {author} {\bibfnamefont {M.~E.}\ \bibnamefont {Cates}},\
  }\bibfield  {title} {\bibinfo {title} {Scalar $\phi$4 field theory for
  active-particle phase separation},\ }\href
  {https://doi.org/10.1038/ncomms5351} {\bibfield  {journal} {\bibinfo
  {journal} {Nature Communications}\ }\textbf {\bibinfo {volume} {5}},\
  \bibinfo {pages} {4351} (\bibinfo {year} {2014})}\BibitemShut {NoStop}%
\bibitem [{\citenamefont {Zwicker}\ \emph {et~al.}(2017)\citenamefont
  {Zwicker}, \citenamefont {Seyboldt}, \citenamefont {Weber}, \citenamefont
  {Hyman},\ and\ \citenamefont {J{\"u}licher}}]{Rabea-Natphy-2017}%
  \BibitemOpen
  \bibfield  {author} {\bibinfo {author} {\bibfnamefont {D.}~\bibnamefont
  {Zwicker}}, \bibinfo {author} {\bibfnamefont {R.}~\bibnamefont {Seyboldt}},
  \bibinfo {author} {\bibfnamefont {C.~A.}\ \bibnamefont {Weber}}, \bibinfo
  {author} {\bibfnamefont {A.~A.}\ \bibnamefont {Hyman}},\ and\ \bibinfo
  {author} {\bibfnamefont {F.}~\bibnamefont {J{\"u}licher}},\ }\bibfield
  {title} {\bibinfo {title} {Growth and division of active droplets provides a
  model for protocells},\ }\href@noop {} {\bibfield  {journal} {\bibinfo
  {journal} {Nat. Phys.}\ }\textbf {\bibinfo {volume} {13}},\ \bibinfo {pages}
  {408} (\bibinfo {year} {2017})}\BibitemShut {NoStop}%
\bibitem [{\citenamefont {Brauns}\ \emph {et~al.}(2020)\citenamefont {Brauns},
  \citenamefont {Halatek},\ and\ \citenamefont
  {Frey}}]{Brauns_Frey_PhysRevX.10.041036}%
  \BibitemOpen
  \bibfield  {author} {\bibinfo {author} {\bibfnamefont {F.}~\bibnamefont
  {Brauns}}, \bibinfo {author} {\bibfnamefont {J.}~\bibnamefont {Halatek}},\
  and\ \bibinfo {author} {\bibfnamefont {E.}~\bibnamefont {Frey}},\ }\bibfield
  {title} {\bibinfo {title} {Phase-space geometry of mass-conserving
  reaction-diffusion dynamics},\ }\href
  {https://doi.org/10.1103/PhysRevX.10.041036} {\bibfield  {journal} {\bibinfo
  {journal} {Phys. Rev. X}\ }\textbf {\bibinfo {volume} {10}},\ \bibinfo
  {pages} {041036} (\bibinfo {year} {2020})}\BibitemShut {NoStop}%
\bibitem [{\citenamefont {Saha}\ \emph {et~al.}(2020)\citenamefont {Saha},
  \citenamefont {{Agudo-Canalejo}},\ and\ \citenamefont
  {Golestanian}}]{saha2020}%
  \BibitemOpen
  \bibfield  {author} {\bibinfo {author} {\bibfnamefont {S.}~\bibnamefont
  {Saha}}, \bibinfo {author} {\bibfnamefont {J.}~\bibnamefont
  {{Agudo-Canalejo}}},\ and\ \bibinfo {author} {\bibfnamefont {R.}~\bibnamefont
  {Golestanian}},\ }\bibfield  {title} {\bibinfo {title} {Scalar {{Active
  Mixtures}}: {{The Nonreciprocal Cahn-Hilliard Model}}},\ }\href
  {https://doi.org/10.1103/PhysRevX.10.041009} {\bibfield  {journal} {\bibinfo
  {journal} {Phys. Rev. X}\ }\textbf {\bibinfo {volume} {10}},\ \bibinfo
  {pages} {041009} (\bibinfo {year} {2020})}\BibitemShut {NoStop}%
\bibitem [{\citenamefont {You}\ \emph {et~al.}(2020{\natexlab{b}})\citenamefont
  {You}, \citenamefont {Baskaran},\ and\ \citenamefont
  {Marchetti}}]{Aparna-non-reciprocity-PNAS-20}%
  \BibitemOpen
  \bibfield  {author} {\bibinfo {author} {\bibfnamefont {Z.}~\bibnamefont
  {You}}, \bibinfo {author} {\bibfnamefont {A.}~\bibnamefont {Baskaran}},\ and\
  \bibinfo {author} {\bibfnamefont {M.~C.}\ \bibnamefont {Marchetti}},\
  }\bibfield  {title} {\bibinfo {title} {Nonreciprocity as a generic route to
  traveling states},\ }\href
  {https://www.pnas.org/doi/abs/10.1073/pnas.2010318117} {\bibfield  {journal}
  {\bibinfo  {journal} {Proc. Nat. Acad. Sci. USA}\ }\textbf {\bibinfo {volume}
  {117}},\ \bibinfo {pages} {197767} (\bibinfo {year}
  {2020}{\natexlab{b}})}\BibitemShut {NoStop}%
\bibitem [{\citenamefont {Soto}\ and\ \citenamefont
  {Golestanian}(2014)}]{soto2014}%
  \BibitemOpen
  \bibfield  {author} {\bibinfo {author} {\bibfnamefont {R.}~\bibnamefont
  {Soto}}\ and\ \bibinfo {author} {\bibfnamefont {R.}~\bibnamefont
  {Golestanian}},\ }\bibfield  {title} {\bibinfo {title} {Self-{{Assembly}} of
  {{Catalytically Active Colloidal Molecules}}: {{Tailoring Activity Through
  Surface Chemistry}}},\ }\href
  {https://doi.org/10.1103/PhysRevLett.112.068301} {\bibfield  {journal}
  {\bibinfo  {journal} {Phys. Rev. Lett.}\ }\textbf {\bibinfo {volume} {112}},\
  \bibinfo {pages} {068301} (\bibinfo {year} {2014})}\BibitemShut {NoStop}%
\bibitem [{\citenamefont {Soto}\ and\ \citenamefont
  {Golestanian}(2015)}]{soto2015}%
  \BibitemOpen
  \bibfield  {author} {\bibinfo {author} {\bibfnamefont {R.}~\bibnamefont
  {Soto}}\ and\ \bibinfo {author} {\bibfnamefont {R.}~\bibnamefont
  {Golestanian}},\ }\bibfield  {title} {\bibinfo {title} {Self-assembly of
  active colloidal molecules with dynamic function},\ }\href
  {https://doi.org/10.1103/PhysRevE.91.052304} {\bibfield  {journal} {\bibinfo
  {journal} {Phys. Rev. E}\ }\textbf {\bibinfo {volume} {91}},\ \bibinfo
  {pages} {052304} (\bibinfo {year} {2015})}\BibitemShut {NoStop}%
\bibitem [{\citenamefont {Saha}\ \emph {et~al.}(2019)\citenamefont {Saha},
  \citenamefont {Ramaswamy},\ and\ \citenamefont {Golestanian}}]{saha2019}%
  \BibitemOpen
  \bibfield  {author} {\bibinfo {author} {\bibfnamefont {S.}~\bibnamefont
  {Saha}}, \bibinfo {author} {\bibfnamefont {S.}~\bibnamefont {Ramaswamy}},\
  and\ \bibinfo {author} {\bibfnamefont {R.}~\bibnamefont {Golestanian}},\
  }\bibfield  {title} {\bibinfo {title} {Pairing, waltzing and scattering of
  chemotactic active colloids},\ }\href
  {https://doi.org/10.1088/1367-2630/ab20fd} {\bibfield  {journal} {\bibinfo
  {journal} {New J. Phys.}\ }\textbf {\bibinfo {volume} {21}},\ \bibinfo
  {pages} {063006} (\bibinfo {year} {2019})}\BibitemShut {NoStop}%
\bibitem [{\citenamefont {Ouazan-Reboul}\ \emph {et~al.}(2023)\citenamefont
  {Ouazan-Reboul}, \citenamefont {Agudo-Canalejo},\ and\ \citenamefont
  {Golestanian}}]{OuazanReboul2023}%
  \BibitemOpen
  \bibfield  {author} {\bibinfo {author} {\bibfnamefont {V.}~\bibnamefont
  {Ouazan-Reboul}}, \bibinfo {author} {\bibfnamefont {J.}~\bibnamefont
  {Agudo-Canalejo}},\ and\ \bibinfo {author} {\bibfnamefont {R.}~\bibnamefont
  {Golestanian}},\ }\bibfield  {title} {\bibinfo {title} {Self-organization of
  primitive metabolic cycles due to non-reciprocal interactions},\ }\href
  {http://dx.doi.org/10.1038/s41467-023-40241-w} {\bibfield  {journal}
  {\bibinfo  {journal} {Nat. Commun.}\ }\textbf {\bibinfo {volume} {14}},\
  \bibinfo {pages} {4496} (\bibinfo {year} {2023})}\BibitemShut {NoStop}%
\bibitem [{\citenamefont {Parkavousi}\ \emph {et~al.}(2025)\citenamefont
  {Parkavousi}, \citenamefont {Rana}, \citenamefont {Golestanian},\ and\
  \citenamefont {Saha}}]{parkavousi2024enhanced}%
  \BibitemOpen
  \bibfield  {author} {\bibinfo {author} {\bibfnamefont {L.}~\bibnamefont
  {Parkavousi}}, \bibinfo {author} {\bibfnamefont {N.}~\bibnamefont {Rana}},
  \bibinfo {author} {\bibfnamefont {R.}~\bibnamefont {Golestanian}},\ and\
  \bibinfo {author} {\bibfnamefont {S.}~\bibnamefont {Saha}},\ }\bibfield
  {title} {\bibinfo {title} {Enhanced stability and chaotic condensates in
  multispecies nonreciprocal mixtures},\ }\href
  {https://doi.org/10.1103/PhysRevLett.134.148301} {\bibfield  {journal}
  {\bibinfo  {journal} {Phys. Rev. Lett.}\ }\textbf {\bibinfo {volume} {134}},\
  \bibinfo {pages} {148301} (\bibinfo {year} {2025})}\BibitemShut {NoStop}%
\bibitem [{\citenamefont {Dinelli}\ \emph {et~al.}(2023)\citenamefont
  {Dinelli}, \citenamefont {O’Byrne}, \citenamefont {Curatolo}, \citenamefont
  {Zhao}, \citenamefont {Sollich},\ and\ \citenamefont
  {Tailleur}}]{dinelli2023non}%
  \BibitemOpen
  \bibfield  {author} {\bibinfo {author} {\bibfnamefont {A.}~\bibnamefont
  {Dinelli}}, \bibinfo {author} {\bibfnamefont {J.}~\bibnamefont {O’Byrne}},
  \bibinfo {author} {\bibfnamefont {A.}~\bibnamefont {Curatolo}}, \bibinfo
  {author} {\bibfnamefont {Y.}~\bibnamefont {Zhao}}, \bibinfo {author}
  {\bibfnamefont {P.}~\bibnamefont {Sollich}},\ and\ \bibinfo {author}
  {\bibfnamefont {J.}~\bibnamefont {Tailleur}},\ }\bibfield  {title} {\bibinfo
  {title} {Non-reciprocity across scales in active mixtures},\ }\href@noop {}
  {\bibfield  {journal} {\bibinfo  {journal} {Nature Communications}\ }\textbf
  {\bibinfo {volume} {14}},\ \bibinfo {pages} {7035} (\bibinfo {year}
  {2023})}\BibitemShut {NoStop}%
\bibitem [{\citenamefont {Loos}\ and\ \citenamefont
  {Klapp}(2020)}]{loos2020irreversibility}%
  \BibitemOpen
  \bibfield  {author} {\bibinfo {author} {\bibfnamefont {S.~A.}\ \bibnamefont
  {Loos}}\ and\ \bibinfo {author} {\bibfnamefont {S.~H.}\ \bibnamefont
  {Klapp}},\ }\bibfield  {title} {\bibinfo {title} {Irreversibility, heat and
  information flows induced by non-reciprocal interactions},\ }\href@noop {}
  {\bibfield  {journal} {\bibinfo  {journal} {New Journal of Physics}\ }\textbf
  {\bibinfo {volume} {22}},\ \bibinfo {pages} {123051} (\bibinfo {year}
  {2020})}\BibitemShut {NoStop}%
\bibitem [{\citenamefont {Fruchart}\ \emph {et~al.}(2021)\citenamefont
  {Fruchart}, \citenamefont {Hanai}, \citenamefont {Littlewood},\ and\
  \citenamefont {Vitelli}}]{fruchart2021}%
  \BibitemOpen
  \bibfield  {author} {\bibinfo {author} {\bibfnamefont {M.}~\bibnamefont
  {Fruchart}}, \bibinfo {author} {\bibfnamefont {R.}~\bibnamefont {Hanai}},
  \bibinfo {author} {\bibfnamefont {P.~B.}\ \bibnamefont {Littlewood}},\ and\
  \bibinfo {author} {\bibfnamefont {V.}~\bibnamefont {Vitelli}},\ }\bibfield
  {title} {\bibinfo {title} {Non-reciprocal phase transitions},\ }\href
  {https://doi.org/10/gjtsw2} {\bibfield  {journal} {\bibinfo  {journal}
  {Nature}\ }\textbf {\bibinfo {volume} {592}},\ \bibinfo {pages} {363}
  (\bibinfo {year} {2021})}\BibitemShut {NoStop}%
\bibitem [{\citenamefont {Duan}\ \emph {et~al.}(2023)\citenamefont {Duan},
  \citenamefont {Agudo-Canalejo}, \citenamefont {Golestanian},\ and\
  \citenamefont {Mahault}}]{BenoitYu2023}%
  \BibitemOpen
  \bibfield  {author} {\bibinfo {author} {\bibfnamefont {Y.}~\bibnamefont
  {Duan}}, \bibinfo {author} {\bibfnamefont {J.}~\bibnamefont
  {Agudo-Canalejo}}, \bibinfo {author} {\bibfnamefont {R.}~\bibnamefont
  {Golestanian}},\ and\ \bibinfo {author} {\bibfnamefont {B.}~\bibnamefont
  {Mahault}},\ }\bibfield  {title} {\bibinfo {title} {Dynamical pattern
  formation without self-attraction in quorum-sensing active matter: The
  interplay between nonreciprocity and motility},\ }\href
  {https://doi.org/10.1103/PhysRevLett.131.148301} {\bibfield  {journal}
  {\bibinfo  {journal} {Phys. Rev. Lett.}\ }\textbf {\bibinfo {volume} {131}},\
  \bibinfo {pages} {148301} (\bibinfo {year} {2023})}\BibitemShut {NoStop}%
\bibitem [{\citenamefont {Tucci}\ \emph {et~al.}(2024)\citenamefont {Tucci},
  \citenamefont {Golestanian},\ and\ \citenamefont
  {Saha}}]{tucci2024nonreciprocal}%
  \BibitemOpen
  \bibfield  {author} {\bibinfo {author} {\bibfnamefont {G.}~\bibnamefont
  {Tucci}}, \bibinfo {author} {\bibfnamefont {R.}~\bibnamefont {Golestanian}},\
  and\ \bibinfo {author} {\bibfnamefont {S.}~\bibnamefont {Saha}},\ }\bibfield
  {title} {\bibinfo {title} {Nonreciprocal collective dynamics in a mixture of
  phoretic janus colloids},\ }\href@noop {} {\bibfield  {journal} {\bibinfo
  {journal} {New J. Phys.}\ }\textbf {\bibinfo {volume} {26}},\ \bibinfo
  {pages} {073006} (\bibinfo {year} {2024})}\BibitemShut {NoStop}%
\bibitem [{\citenamefont {Chen}\ \emph {et~al.}(2024)\citenamefont {Chen},
  \citenamefont {Lei}, \citenamefont {Xiang}, \citenamefont {Duan},
  \citenamefont {Peng},\ and\ \citenamefont {Zhang}}]{chen2024emergent}%
  \BibitemOpen
  \bibfield  {author} {\bibinfo {author} {\bibfnamefont {J.}~\bibnamefont
  {Chen}}, \bibinfo {author} {\bibfnamefont {X.}~\bibnamefont {Lei}}, \bibinfo
  {author} {\bibfnamefont {Y.}~\bibnamefont {Xiang}}, \bibinfo {author}
  {\bibfnamefont {M.}~\bibnamefont {Duan}}, \bibinfo {author} {\bibfnamefont
  {X.}~\bibnamefont {Peng}},\ and\ \bibinfo {author} {\bibfnamefont
  {H.}~\bibnamefont {Zhang}},\ }\bibfield  {title} {\bibinfo {title} {Emergent
  chirality and hyperuniformity in an active mixture with nonreciprocal
  interactions},\ }\href@noop {} {\bibfield  {journal} {\bibinfo  {journal}
  {Physical Review Letters}\ }\textbf {\bibinfo {volume} {132}},\ \bibinfo
  {pages} {118301} (\bibinfo {year} {2024})}\BibitemShut {NoStop}%
\bibitem [{\citenamefont {Guillet}\ \emph {et~al.}(2025)\citenamefont
  {Guillet}, \citenamefont {Poncet}, \citenamefont {Le~Blay}, \citenamefont
  {Irvine}, \citenamefont {Vitelli},\ and\ \citenamefont
  {Bartolo}}]{guillet2025melting}%
  \BibitemOpen
  \bibfield  {author} {\bibinfo {author} {\bibfnamefont {S.}~\bibnamefont
  {Guillet}}, \bibinfo {author} {\bibfnamefont {A.}~\bibnamefont {Poncet}},
  \bibinfo {author} {\bibfnamefont {M.}~\bibnamefont {Le~Blay}}, \bibinfo
  {author} {\bibfnamefont {W.~T.}\ \bibnamefont {Irvine}}, \bibinfo {author}
  {\bibfnamefont {V.}~\bibnamefont {Vitelli}},\ and\ \bibinfo {author}
  {\bibfnamefont {D.}~\bibnamefont {Bartolo}},\ }\bibfield  {title} {\bibinfo
  {title} {Melting of nonreciprocal solids: How dislocations propel and fission
  in flowing crystals},\ }\href@noop {} {\bibfield  {journal} {\bibinfo
  {journal} {Proceedings of the National Academy of Sciences}\ }\textbf
  {\bibinfo {volume} {122}},\ \bibinfo {pages} {e2412993122} (\bibinfo {year}
  {2025})}\BibitemShut {NoStop}%
\bibitem [{\citenamefont {Tan}\ \emph {et~al.}(2022)\citenamefont {Tan},
  \citenamefont {Mietke}, \citenamefont {Li}, \citenamefont {Chen},
  \citenamefont {Higinbotham}, \citenamefont {Foster}, \citenamefont {Gokhale},
  \citenamefont {Dunkel},\ and\ \citenamefont {Fakhri}}]{tan2022odd}%
  \BibitemOpen
  \bibfield  {author} {\bibinfo {author} {\bibfnamefont {T.~H.}\ \bibnamefont
  {Tan}}, \bibinfo {author} {\bibfnamefont {A.}~\bibnamefont {Mietke}},
  \bibinfo {author} {\bibfnamefont {J.}~\bibnamefont {Li}}, \bibinfo {author}
  {\bibfnamefont {Y.}~\bibnamefont {Chen}}, \bibinfo {author} {\bibfnamefont
  {H.}~\bibnamefont {Higinbotham}}, \bibinfo {author} {\bibfnamefont {P.~J.}\
  \bibnamefont {Foster}}, \bibinfo {author} {\bibfnamefont {S.}~\bibnamefont
  {Gokhale}}, \bibinfo {author} {\bibfnamefont {J.}~\bibnamefont {Dunkel}},\
  and\ \bibinfo {author} {\bibfnamefont {N.}~\bibnamefont {Fakhri}},\
  }\bibfield  {title} {\bibinfo {title} {Odd dynamics of living chiral
  crystals},\ }\href@noop {} {\bibfield  {journal} {\bibinfo  {journal}
  {Nature}\ }\textbf {\bibinfo {volume} {607}},\ \bibinfo {pages} {287}
  (\bibinfo {year} {2022})}\BibitemShut {NoStop}%
\bibitem [{\citenamefont {Chao}\ \emph {et~al.}(2026)\citenamefont {Chao},
  \citenamefont {Gokhale}, \citenamefont {Lin}, \citenamefont {Hastewell},
  \citenamefont {Bacanu}, \citenamefont {Chen}, \citenamefont {Li},
  \citenamefont {Liu}, \citenamefont {Lee}, \citenamefont {Dunkel} \emph
  {et~al.}}]{chao2026selective}%
  \BibitemOpen
  \bibfield  {author} {\bibinfo {author} {\bibfnamefont {Y.-C.}\ \bibnamefont
  {Chao}}, \bibinfo {author} {\bibfnamefont {S.}~\bibnamefont {Gokhale}},
  \bibinfo {author} {\bibfnamefont {L.}~\bibnamefont {Lin}}, \bibinfo {author}
  {\bibfnamefont {A.}~\bibnamefont {Hastewell}}, \bibinfo {author}
  {\bibfnamefont {A.}~\bibnamefont {Bacanu}}, \bibinfo {author} {\bibfnamefont
  {Y.}~\bibnamefont {Chen}}, \bibinfo {author} {\bibfnamefont {J.}~\bibnamefont
  {Li}}, \bibinfo {author} {\bibfnamefont {J.}~\bibnamefont {Liu}}, \bibinfo
  {author} {\bibfnamefont {H.}~\bibnamefont {Lee}}, \bibinfo {author}
  {\bibfnamefont {J.}~\bibnamefont {Dunkel}}, \emph {et~al.},\ }\bibfield
  {title} {\bibinfo {title} {Selective excitation of work-generating cycles in
  non-reciprocal living solids},\ }\href@noop {} {\bibfield  {journal}
  {\bibinfo  {journal} {Nature Physics}\ ,\ \bibinfo {pages} {1}} (\bibinfo
  {year} {2026})}\BibitemShut {NoStop}%
\bibitem [{\citenamefont {Kant}\ \emph {et~al.}(2024)\citenamefont {Kant},
  \citenamefont {Gupta}, \citenamefont {Soni}, \citenamefont {Sood},\ and\
  \citenamefont {Ramaswamy}}]{SriramExptNRCH_PhysRevLett.133.208301}%
  \BibitemOpen
  \bibfield  {author} {\bibinfo {author} {\bibfnamefont {R.}~\bibnamefont
  {Kant}}, \bibinfo {author} {\bibfnamefont {R.~K.}\ \bibnamefont {Gupta}},
  \bibinfo {author} {\bibfnamefont {H.}~\bibnamefont {Soni}}, \bibinfo {author}
  {\bibfnamefont {A.~K.}\ \bibnamefont {Sood}},\ and\ \bibinfo {author}
  {\bibfnamefont {S.}~\bibnamefont {Ramaswamy}},\ }\bibfield  {title} {\bibinfo
  {title} {Bulk condensation by an active interface},\ }\href
  {https://doi.org/10.1103/PhysRevLett.133.208301} {\bibfield  {journal}
  {\bibinfo  {journal} {Phys. Rev. Lett.}\ }\textbf {\bibinfo {volume} {133}},\
  \bibinfo {pages} {208301} (\bibinfo {year} {2024})}\BibitemShut {NoStop}%
\bibitem [{\citenamefont {Saha}\ and\ \citenamefont
  {Golestanian}(2022)}]{saha2022}%
  \BibitemOpen
  \bibfield  {author} {\bibinfo {author} {\bibfnamefont {S.}~\bibnamefont
  {Saha}}\ and\ \bibinfo {author} {\bibfnamefont {R.}~\bibnamefont
  {Golestanian}},\ }\href {https://doi.org/10.48550/arXiv.2208.14985} {\bibinfo
  {title} {Effervescent waves in a binary mixture with non-reciprocal
  couplings}} (\bibinfo {year} {2022}),\ \Eprint
  {https://arxiv.org/abs/2208.14985} {2208.14985} \BibitemShut {NoStop}%
\bibitem [{\citenamefont {Pisegna}\ \emph {et~al.}(2024)\citenamefont
  {Pisegna}, \citenamefont {Saha},\ and\ \citenamefont
  {Golestanian}}]{Pisegna2024}%
  \BibitemOpen
  \bibfield  {author} {\bibinfo {author} {\bibfnamefont {G.}~\bibnamefont
  {Pisegna}}, \bibinfo {author} {\bibfnamefont {S.}~\bibnamefont {Saha}},\ and\
  \bibinfo {author} {\bibfnamefont {R.}~\bibnamefont {Golestanian}},\
  }\bibfield  {title} {\bibinfo {title} {Emergent polar order in nonpolar
  mixtures with nonreciprocal interactions},\ }\href
  {https://doi.org/10.1073/pnas.2407705121} {\bibfield  {journal} {\bibinfo
  {journal} {Proceedings of the National Academy of Sciences}\ }\textbf
  {\bibinfo {volume} {121}},\ \bibinfo {pages} {e2407705121} (\bibinfo {year}
  {2024})}\BibitemShut {NoStop}%
\bibitem [{\citenamefont
  {Saha}(2024)}]{saha2024phasecoexistencenonreciprocalcahnhilliard}%
  \BibitemOpen
  \bibfield  {author} {\bibinfo {author} {\bibfnamefont {S.}~\bibnamefont
  {Saha}},\ }\href {https://arxiv.org/abs/2402.10057} {\bibinfo {title} {Phase
  coexistence in the non-reciprocal cahn-hilliard model}} (\bibinfo {year}
  {2024}),\ \Eprint {https://arxiv.org/abs/2402.10057} {arXiv:2402.10057
  [cond-mat.soft]} \BibitemShut {NoStop}%
\bibitem [{\citenamefont {Greve}\ \emph {et~al.}(2025)\citenamefont {Greve},
  \citenamefont {Lovato}, \citenamefont {Frohoff-H\"ulsmann},\ and\
  \citenamefont {Thiele}}]{ThielePS}%
  \BibitemOpen
  \bibfield  {author} {\bibinfo {author} {\bibfnamefont {D.}~\bibnamefont
  {Greve}}, \bibinfo {author} {\bibfnamefont {G.}~\bibnamefont {Lovato}},
  \bibinfo {author} {\bibfnamefont {T.}~\bibnamefont {Frohoff-H\"ulsmann}},\
  and\ \bibinfo {author} {\bibfnamefont {U.}~\bibnamefont {Thiele}},\
  }\bibfield  {title} {\bibinfo {title} {Coexistence of uniform and oscillatory
  states resulting from nonreciprocity and conservation laws},\ }\href
  {https://doi.org/10.1103/PhysRevLett.134.018303} {\bibfield  {journal}
  {\bibinfo  {journal} {Phys. Rev. Lett.}\ }\textbf {\bibinfo {volume} {134}},\
  \bibinfo {pages} {018303} (\bibinfo {year} {2025})}\BibitemShut {NoStop}%
\bibitem [{\citenamefont {McCall}\ \emph {et~al.}(2025)\citenamefont {McCall},
  \citenamefont {Kim}, \citenamefont {Shevchenko}, \citenamefont
  {Ruer-Gru{\ss}}, \citenamefont {Peychl}, \citenamefont {Guck}, \citenamefont
  {Shevchenko}, \citenamefont {Hyman},\ and\ \citenamefont
  {Brugu{\'e}s}}]{McCall2025}%
  \BibitemOpen
  \bibfield  {author} {\bibinfo {author} {\bibfnamefont {P.~M.}\ \bibnamefont
  {McCall}}, \bibinfo {author} {\bibfnamefont {K.}~\bibnamefont {Kim}},
  \bibinfo {author} {\bibfnamefont {A.}~\bibnamefont {Shevchenko}}, \bibinfo
  {author} {\bibfnamefont {M.}~\bibnamefont {Ruer-Gru{\ss}}}, \bibinfo {author}
  {\bibfnamefont {J.}~\bibnamefont {Peychl}}, \bibinfo {author} {\bibfnamefont
  {J.}~\bibnamefont {Guck}}, \bibinfo {author} {\bibfnamefont {A.}~\bibnamefont
  {Shevchenko}}, \bibinfo {author} {\bibfnamefont {A.~A.}\ \bibnamefont
  {Hyman}},\ and\ \bibinfo {author} {\bibfnamefont {J.}~\bibnamefont
  {Brugu{\'e}s}},\ }\bibfield  {title} {\bibinfo {title} {A label-free method
  for measuring the composition of multicomponent biomolecular condensates},\
  }\bibfield  {journal} {\bibinfo  {journal} {Nature Chemistry}\ }\href
  {https://doi.org/10.1038/s41557-025-01928-3} {10.1038/s41557-025-01928-3}
  (\bibinfo {year} {2025})\BibitemShut {NoStop}%
\bibitem [{\citenamefont {Klein}\ \emph {et~al.}(2020)\citenamefont {Klein},
  \citenamefont {Boija}, \citenamefont {Afeyan}, \citenamefont {Hawken},
  \citenamefont {Fan}, \citenamefont {Dall'Agnese}, \citenamefont {Oksuz},
  \citenamefont {Henninger}, \citenamefont {Shrinivas}, \citenamefont {Sabari},
  \citenamefont {Sagi}, \citenamefont {Clark}, \citenamefont {Platt},
  \citenamefont {Kar}, \citenamefont {McCall}, \citenamefont {Zamudio},
  \citenamefont {Manteiga}, \citenamefont {Coffey}, \citenamefont {Li},
  \citenamefont {Hannett}, \citenamefont {Guo}, \citenamefont {Decker},
  \citenamefont {Lee}, \citenamefont {Zhang}, \citenamefont {Weng},
  \citenamefont {Taatjes}, \citenamefont {Chakraborty}, \citenamefont {Sharp},
  \citenamefont {Chang}, \citenamefont {Hyman}, \citenamefont {Gray},\ and\
  \citenamefont {Young}}]{Klein2020}%
  \BibitemOpen
  \bibfield  {author} {\bibinfo {author} {\bibfnamefont {I.~A.}\ \bibnamefont
  {Klein}}, \bibinfo {author} {\bibfnamefont {A.}~\bibnamefont {Boija}},
  \bibinfo {author} {\bibfnamefont {L.~K.}\ \bibnamefont {Afeyan}}, \bibinfo
  {author} {\bibfnamefont {S.~W.}\ \bibnamefont {Hawken}}, \bibinfo {author}
  {\bibfnamefont {M.}~\bibnamefont {Fan}}, \bibinfo {author} {\bibfnamefont
  {A.}~\bibnamefont {Dall'Agnese}}, \bibinfo {author} {\bibfnamefont
  {O.}~\bibnamefont {Oksuz}}, \bibinfo {author} {\bibfnamefont {J.~E.}\
  \bibnamefont {Henninger}}, \bibinfo {author} {\bibfnamefont {K.}~\bibnamefont
  {Shrinivas}}, \bibinfo {author} {\bibfnamefont {B.~R.}\ \bibnamefont
  {Sabari}}, \bibinfo {author} {\bibfnamefont {I.}~\bibnamefont {Sagi}},
  \bibinfo {author} {\bibfnamefont {V.~E.}\ \bibnamefont {Clark}}, \bibinfo
  {author} {\bibfnamefont {J.~M.}\ \bibnamefont {Platt}}, \bibinfo {author}
  {\bibfnamefont {M.}~\bibnamefont {Kar}}, \bibinfo {author} {\bibfnamefont
  {P.~M.}\ \bibnamefont {McCall}}, \bibinfo {author} {\bibfnamefont {A.~V.}\
  \bibnamefont {Zamudio}}, \bibinfo {author} {\bibfnamefont {J.~C.}\
  \bibnamefont {Manteiga}}, \bibinfo {author} {\bibfnamefont {E.~L.}\
  \bibnamefont {Coffey}}, \bibinfo {author} {\bibfnamefont {C.~H.}\
  \bibnamefont {Li}}, \bibinfo {author} {\bibfnamefont {N.~M.}\ \bibnamefont
  {Hannett}}, \bibinfo {author} {\bibfnamefont {Y.~E.}\ \bibnamefont {Guo}},
  \bibinfo {author} {\bibfnamefont {T.-M.}\ \bibnamefont {Decker}}, \bibinfo
  {author} {\bibfnamefont {T.~I.}\ \bibnamefont {Lee}}, \bibinfo {author}
  {\bibfnamefont {T.}~\bibnamefont {Zhang}}, \bibinfo {author} {\bibfnamefont
  {J.-K.}\ \bibnamefont {Weng}}, \bibinfo {author} {\bibfnamefont {D.~J.}\
  \bibnamefont {Taatjes}}, \bibinfo {author} {\bibfnamefont {A.}~\bibnamefont
  {Chakraborty}}, \bibinfo {author} {\bibfnamefont {P.~A.}\ \bibnamefont
  {Sharp}}, \bibinfo {author} {\bibfnamefont {Y.~T.}\ \bibnamefont {Chang}},
  \bibinfo {author} {\bibfnamefont {A.~A.}\ \bibnamefont {Hyman}}, \bibinfo
  {author} {\bibfnamefont {N.~S.}\ \bibnamefont {Gray}},\ and\ \bibinfo
  {author} {\bibfnamefont {R.~A.}\ \bibnamefont {Young}},\ }\bibfield  {title}
  {\bibinfo {title} {Partitioning of cancer therapeutics in nuclear
  condensates},\ }\href {https://doi.org/10.1126/science.aaz4427} {\bibfield
  {journal} {\bibinfo  {journal} {Science}\ }\textbf {\bibinfo {volume}
  {368}},\ \bibinfo {pages} {1386} (\bibinfo {year} {2020})}\BibitemShut
  {NoStop}%
\bibitem [{\citenamefont {Chaderjian}\ \emph {et~al.}(2025)\citenamefont
  {Chaderjian}, \citenamefont {Wilken},\ and\ \citenamefont
  {Saleh}}]{chaderjian2025diversedistinctdenselypacked}%
  \BibitemOpen
  \bibfield  {author} {\bibinfo {author} {\bibfnamefont {A.~S.}\ \bibnamefont
  {Chaderjian}}, \bibinfo {author} {\bibfnamefont {S.}~\bibnamefont {Wilken}},\
  and\ \bibinfo {author} {\bibfnamefont {O.~A.}\ \bibnamefont {Saleh}},\ }\href
  {https://arxiv.org/abs/2508.18574} {\bibinfo {title} {Diverse, distinct, and
  densely packed dna droplets}} (\bibinfo {year} {2025}),\ \Eprint
  {https://arxiv.org/abs/2508.18574} {arXiv:2508.18574 [cond-mat.soft]}
  \BibitemShut {NoStop}%
\bibitem [{\citenamefont {Schehr}\ \emph {et~al.}(2017)\citenamefont {Schehr},
  \citenamefont {Altland}, \citenamefont {Fyodorov}, \citenamefont
  {O'Connell},\ and\ \citenamefont {Cugliandolo}}]{schehr2017}%
  \BibitemOpen
  \bibinfo {editor} {\bibfnamefont {G.}~\bibnamefont {Schehr}}, \bibinfo
  {editor} {\bibfnamefont {A.}~\bibnamefont {Altland}}, \bibinfo {editor}
  {\bibfnamefont {Y.~V.}\ \bibnamefont {Fyodorov}}, \bibinfo {editor}
  {\bibfnamefont {N.}~\bibnamefont {O'Connell}},\ and\ \bibinfo {editor}
  {\bibfnamefont {L.~F.}\ \bibnamefont {Cugliandolo}},\ eds.,\ \href@noop {}
  {\emph {\bibinfo {title} {Stochastic Processes and Random Matrices: Lecture
  Notes of the {{Les Houches Summer School}}: Volume 104, 6th-31st {{July}}
  2015}}},\ \bibinfo {edition} {first edition}\ ed.,\ \bibinfo {series}
  {Lecture Notes of the {{Les Houches Summer School}}}\ No.\ \bibinfo {number}
  {volume 104}\ (\bibinfo  {publisher} {{Oxford University Press}},\ \bibinfo
  {address} {{Oxford, United Kingdom}},\ \bibinfo {year} {2017})\BibitemShut
  {NoStop}%
\bibitem [{\citenamefont {Mehta}(2004)}]{mehta2004}%
  \BibitemOpen
  \bibfield  {author} {\bibinfo {author} {\bibfnamefont {M.~L.}\ \bibnamefont
  {Mehta}},\ }\href {http://www.myilibrary.com?id=96818} {\emph {\bibinfo
  {title} {Random Matrices}}},\ \bibinfo {edition} {3rd}\ ed.,\ \bibinfo
  {series} {Pure and Applied Mathematics}\ No.\ \bibinfo {number} {v. 142}\
  (\bibinfo  {publisher} {{Academic Press}},\ \bibinfo {address} {{Amsterdam ;
  San Diego, CA}},\ \bibinfo {year} {2004})\BibitemShut {NoStop}%
\bibitem [{\citenamefont {Bordenave}\ and\ \citenamefont
  {Chafa{\"\i}}(2012)}]{bordenave2012around}%
  \BibitemOpen
  \bibfield  {author} {\bibinfo {author} {\bibfnamefont {C.}~\bibnamefont
  {Bordenave}}\ and\ \bibinfo {author} {\bibfnamefont {D.}~\bibnamefont
  {Chafa{\"\i}}},\ }\bibfield  {title} {\bibinfo {title} {Around the circular
  law},\ }\href@noop {} {\  (\bibinfo {year} {2012})}\BibitemShut {NoStop}%
\bibitem [{\citenamefont {Bai}\ and\ \citenamefont
  {Silverstein}(2010)}]{bai2010spectral}%
  \BibitemOpen
  \bibfield  {author} {\bibinfo {author} {\bibfnamefont {Z.}~\bibnamefont
  {Bai}}\ and\ \bibinfo {author} {\bibfnamefont {J.~W.}\ \bibnamefont
  {Silverstein}},\ }\href@noop {} {\emph {\bibinfo {title} {Spectral analysis
  of large dimensional random matrices}}},\ Vol.~\bibinfo {volume} {20}\
  (\bibinfo  {publisher} {Springer},\ \bibinfo {year} {2010})\BibitemShut
  {NoStop}%
\bibitem [{\citenamefont {Wigner}(1958)}]{wigner2}%
  \BibitemOpen
  \bibfield  {author} {\bibinfo {author} {\bibfnamefont {E.~P.}\ \bibnamefont
  {Wigner}},\ }\bibfield  {title} {\bibinfo {title} {On the distribution of the
  roots of certain symmetric matrices},\ }\href
  {http://www.jstor.org/stable/1970008} {\bibfield  {journal} {\bibinfo
  {journal} {Annals of Mathematics}\ }\textbf {\bibinfo {volume} {67}},\
  \bibinfo {pages} {325} (\bibinfo {year} {1958})}\BibitemShut {NoStop}%
\bibitem [{\citenamefont {May}(1972)}]{MAY1972}%
  \BibitemOpen
  \bibfield  {author} {\bibinfo {author} {\bibfnamefont {R.~M.}\ \bibnamefont
  {May}},\ }\bibfield  {title} {\bibinfo {title} {Will a large complex system
  be stable?},\ }\href {https://doi.org/10.1038/238413a0} {\bibfield  {journal}
  {\bibinfo  {journal} {Nature}\ }\textbf {\bibinfo {volume} {238}},\ \bibinfo
  {pages} {413} (\bibinfo {year} {1972})}\BibitemShut {NoStop}%
\bibitem [{\citenamefont {Mougi}\ and\ \citenamefont
  {Kondoh}(2012)}]{mougi2012diversity}%
  \BibitemOpen
  \bibfield  {author} {\bibinfo {author} {\bibfnamefont {A.}~\bibnamefont
  {Mougi}}\ and\ \bibinfo {author} {\bibfnamefont {M.}~\bibnamefont {Kondoh}},\
  }\bibfield  {title} {\bibinfo {title} {Diversity of interaction types and
  ecological community stability},\ }\href@noop {} {\bibfield  {journal}
  {\bibinfo  {journal} {Science}\ }\textbf {\bibinfo {volume} {337}},\ \bibinfo
  {pages} {349} (\bibinfo {year} {2012})}\BibitemShut {NoStop}%
\bibitem [{\citenamefont {Altieri}\ and\ \citenamefont
  {Biroli}(2022)}]{altieri2022effects}%
  \BibitemOpen
  \bibfield  {author} {\bibinfo {author} {\bibfnamefont {A.}~\bibnamefont
  {Altieri}}\ and\ \bibinfo {author} {\bibfnamefont {G.}~\bibnamefont
  {Biroli}},\ }\bibfield  {title} {\bibinfo {title} {Effects of intraspecific
  cooperative interactions in large ecosystems},\ }\href@noop {} {\bibfield
  {journal} {\bibinfo  {journal} {SciPost Physics}\ }\textbf {\bibinfo {volume}
  {12}},\ \bibinfo {pages} {013} (\bibinfo {year} {2022})}\BibitemShut
  {NoStop}%
\bibitem [{\citenamefont {Biroli}\ \emph {et~al.}(2018)\citenamefont {Biroli},
  \citenamefont {Bunin},\ and\ \citenamefont
  {Cammarota}}]{biroli2018marginally}%
  \BibitemOpen
  \bibfield  {author} {\bibinfo {author} {\bibfnamefont {G.}~\bibnamefont
  {Biroli}}, \bibinfo {author} {\bibfnamefont {G.}~\bibnamefont {Bunin}},\ and\
  \bibinfo {author} {\bibfnamefont {C.}~\bibnamefont {Cammarota}},\ }\bibfield
  {title} {\bibinfo {title} {Marginally stable equilibria in critical
  ecosystems},\ }\href@noop {} {\bibfield  {journal} {\bibinfo  {journal} {New
  Journal of Physics}\ }\textbf {\bibinfo {volume} {20}},\ \bibinfo {pages}
  {083051} (\bibinfo {year} {2018})}\BibitemShut {NoStop}%
\bibitem [{\citenamefont {Baron}\ \emph {et~al.}(2022)\citenamefont {Baron},
  \citenamefont {Jewell}, \citenamefont {Ryder},\ and\ \citenamefont
  {Galla}}]{baron2022eigenvalues}%
  \BibitemOpen
  \bibfield  {author} {\bibinfo {author} {\bibfnamefont {J.~W.}\ \bibnamefont
  {Baron}}, \bibinfo {author} {\bibfnamefont {T.~J.}\ \bibnamefont {Jewell}},
  \bibinfo {author} {\bibfnamefont {C.}~\bibnamefont {Ryder}},\ and\ \bibinfo
  {author} {\bibfnamefont {T.}~\bibnamefont {Galla}},\ }\bibfield  {title}
  {\bibinfo {title} {Eigenvalues of random matrices with generalized
  correlations: A path integral approach},\ }\href@noop {} {\bibfield
  {journal} {\bibinfo  {journal} {Physical Review Letters}\ }\textbf {\bibinfo
  {volume} {128}},\ \bibinfo {pages} {120601} (\bibinfo {year}
  {2022})}\BibitemShut {NoStop}%
\bibitem [{\citenamefont {Baron}\ and\ \citenamefont
  {Galla}(2020)}]{baron2020dispersal}%
  \BibitemOpen
  \bibfield  {author} {\bibinfo {author} {\bibfnamefont {J.~W.}\ \bibnamefont
  {Baron}}\ and\ \bibinfo {author} {\bibfnamefont {T.}~\bibnamefont {Galla}},\
  }\bibfield  {title} {\bibinfo {title} {Dispersal-induced instability in
  complex ecosystems},\ }\href@noop {} {\bibfield  {journal} {\bibinfo
  {journal} {Nature communications}\ }\textbf {\bibinfo {volume} {11}},\
  \bibinfo {pages} {6032} (\bibinfo {year} {2020})}\BibitemShut {NoStop}%
\bibitem [{\citenamefont {M{\'e}zard}\ \emph {et~al.}(1987)\citenamefont
  {M{\'e}zard}, \citenamefont {Parisi},\ and\ \citenamefont
  {Virasoro}}]{mezard1987spin}%
  \BibitemOpen
  \bibfield  {author} {\bibinfo {author} {\bibfnamefont {M.}~\bibnamefont
  {M{\'e}zard}}, \bibinfo {author} {\bibfnamefont {G.}~\bibnamefont {Parisi}},\
  and\ \bibinfo {author} {\bibfnamefont {M.~A.}\ \bibnamefont {Virasoro}},\
  }\href@noop {} {\emph {\bibinfo {title} {Spin glass theory and beyond: An
  Introduction to the Replica Method and Its Applications}}},\ Vol.~\bibinfo
  {volume} {9}\ (\bibinfo  {publisher} {World Scientific Publishing Company},\
  \bibinfo {year} {1987})\BibitemShut {NoStop}%
\bibitem [{\citenamefont {Crisanti}\ and\ \citenamefont
  {Sompolinsky}(1988)}]{crisanti1988dynamics}%
  \BibitemOpen
  \bibfield  {author} {\bibinfo {author} {\bibfnamefont {A.}~\bibnamefont
  {Crisanti}}\ and\ \bibinfo {author} {\bibfnamefont {H.}~\bibnamefont
  {Sompolinsky}},\ }\bibfield  {title} {\bibinfo {title} {Dynamics of spin
  systems with randomly asymmetric bonds: Ising spins and glauber dynamics},\
  }\href@noop {} {\bibfield  {journal} {\bibinfo  {journal} {Physical Review
  A}\ }\textbf {\bibinfo {volume} {37}},\ \bibinfo {pages} {4865} (\bibinfo
  {year} {1988})}\BibitemShut {NoStop}%
\bibitem [{\citenamefont {Crisanti}\ and\ \citenamefont
  {Sompolinsky}(1987)}]{crisanti1987dynamics}%
  \BibitemOpen
  \bibfield  {author} {\bibinfo {author} {\bibfnamefont {A.}~\bibnamefont
  {Crisanti}}\ and\ \bibinfo {author} {\bibfnamefont {H.}~\bibnamefont
  {Sompolinsky}},\ }\bibfield  {title} {\bibinfo {title} {Dynamics of spin
  systems with randomly asymmetric bonds: Langevin dynamics and a spherical
  model},\ }\href@noop {} {\bibfield  {journal} {\bibinfo  {journal} {Physical
  Review A}\ }\textbf {\bibinfo {volume} {36}},\ \bibinfo {pages} {4922}
  (\bibinfo {year} {1987})}\BibitemShut {NoStop}%
\bibitem [{\citenamefont {Teixeira}\ \emph {et~al.}(2024)\citenamefont
  {Teixeira}, \citenamefont {Carugno}, \citenamefont {Neri},\ and\
  \citenamefont {Sartori}}]{Liquid_Hopfield}%
  \BibitemOpen
  \bibfield  {author} {\bibinfo {author} {\bibfnamefont {R.~B.}\ \bibnamefont
  {Teixeira}}, \bibinfo {author} {\bibfnamefont {G.}~\bibnamefont {Carugno}},
  \bibinfo {author} {\bibfnamefont {I.}~\bibnamefont {Neri}},\ and\ \bibinfo
  {author} {\bibfnamefont {P.}~\bibnamefont {Sartori}},\ }\bibfield  {title}
  {\bibinfo {title} {Liquid hopfield model: Retrieval and localization in
  multicomponent liquid mixtures},\ }\href
  {https://doi.org/10.1073/pnas.2320504121} {\bibfield  {journal} {\bibinfo
  {journal} {Proceedings of the National Academy of Sciences}\ }\textbf
  {\bibinfo {volume} {121}},\ \bibinfo {pages} {e2320504121} (\bibinfo {year}
  {2024})},\ \Eprint
  {https://arxiv.org/abs/https://www.pnas.org/doi/pdf/10.1073/pnas.2320504121}
  {https://www.pnas.org/doi/pdf/10.1073/pnas.2320504121} \BibitemShut {NoStop}%
\bibitem [{\citenamefont {Dinelli}\ \emph {et~al.}(2025)\citenamefont
  {Dinelli}, \citenamefont {Altieri},\ and\ \citenamefont
  {Tailleur}}]{dinelli2025randommotilityregulationdrives}%
  \BibitemOpen
  \bibfield  {author} {\bibinfo {author} {\bibfnamefont {A.}~\bibnamefont
  {Dinelli}}, \bibinfo {author} {\bibfnamefont {A.}~\bibnamefont {Altieri}},\
  and\ \bibinfo {author} {\bibfnamefont {J.}~\bibnamefont {Tailleur}},\ }\href
  {https://arxiv.org/abs/2503.12692} {\bibinfo {title} {Random motility
  regulation drives the fragmentation of microbial ecosystems}} (\bibinfo
  {year} {2025}),\ \Eprint {https://arxiv.org/abs/2503.12692} {arXiv:2503.12692
  [cond-mat.stat-mech]} \BibitemShut {NoStop}%
\bibitem [{\citenamefont {Thewes}\ \emph {et~al.}(2023)\citenamefont {Thewes},
  \citenamefont {Kr{\"u}ger},\ and\ \citenamefont
  {Sollich}}]{thewes2023composition}%
  \BibitemOpen
  \bibfield  {author} {\bibinfo {author} {\bibfnamefont {F.~C.}\ \bibnamefont
  {Thewes}}, \bibinfo {author} {\bibfnamefont {M.}~\bibnamefont {Kr{\"u}ger}},\
  and\ \bibinfo {author} {\bibfnamefont {P.}~\bibnamefont {Sollich}},\
  }\bibfield  {title} {\bibinfo {title} {Composition dependent instabilities in
  mixtures with many components},\ }\href@noop {} {\bibfield  {journal}
  {\bibinfo  {journal} {Physical Review Letters}\ }\textbf {\bibinfo {volume}
  {131}},\ \bibinfo {pages} {058401} (\bibinfo {year} {2023})}\BibitemShut
  {NoStop}%
\bibitem [{\citenamefont {Chaikin}\ and\ \citenamefont
  {Lubensky}(1995)}]{Chaikin_Lubensky_1995}%
  \BibitemOpen
  \bibfield  {author} {\bibinfo {author} {\bibfnamefont {P.~M.}\ \bibnamefont
  {Chaikin}}\ and\ \bibinfo {author} {\bibfnamefont {T.~C.}\ \bibnamefont
  {Lubensky}},\ }\href@noop {} {\emph {\bibinfo {title} {Principles of
  Condensed Matter Physics}}}\ (\bibinfo  {publisher} {Cambridge University
  Press},\ \bibinfo {year} {1995})\BibitemShut {NoStop}%
\bibitem [{\citenamefont {Frohoff-Hülsmann}\ and\ \citenamefont
  {Thiele}(2021)}]{ThieleTuring}%
  \BibitemOpen
  \bibfield  {author} {\bibinfo {author} {\bibfnamefont {T.}~\bibnamefont
  {Frohoff-Hülsmann}}\ and\ \bibinfo {author} {\bibfnamefont {U.}~\bibnamefont
  {Thiele}},\ }\bibfield  {title} {\bibinfo {title} {{Localized states in
  coupled Cahn–Hilliard equations}},\ }\href
  {https://doi.org/10.1093/imamat/hxab026} {\bibfield  {journal} {\bibinfo
  {journal} {IMA Journal of Applied Mathematics}\ }\textbf {\bibinfo {volume}
  {86}},\ \bibinfo {pages} {924} (\bibinfo {year} {2021})},\ \Eprint
  {https://arxiv.org/abs/https://academic.oup.com/imamat/article-pdf/86/5/924/40744935/hxab026.pdf}
  {https://academic.oup.com/imamat/article-pdf/86/5/924/40744935/hxab026.pdf}
  \BibitemShut {NoStop}%
\bibitem [{\citenamefont {Cure}\ and\ \citenamefont
  {Neri}(2023)}]{cure2023antagonistic}%
  \BibitemOpen
  \bibfield  {author} {\bibinfo {author} {\bibfnamefont {S.}~\bibnamefont
  {Cure}}\ and\ \bibinfo {author} {\bibfnamefont {I.}~\bibnamefont {Neri}},\
  }\bibfield  {title} {\bibinfo {title} {Antagonistic interactions can
  stabilise fixed points in heterogeneous linear dynamical systems},\
  }\href@noop {} {\bibfield  {journal} {\bibinfo  {journal} {SciPost Physics}\
  }\textbf {\bibinfo {volume} {14}},\ \bibinfo {pages} {093} (\bibinfo {year}
  {2023})}\BibitemShut {NoStop}%
\bibitem [{\citenamefont {Matsuzawa}\ \emph {et~al.}(2026)\citenamefont
  {Matsuzawa}, \citenamefont {Varma}, \citenamefont {Bate}, \citenamefont
  {Lorenz}, \citenamefont {Larina}, \citenamefont {Bauermann}, \citenamefont
  {Matthias}, \citenamefont {Grubic}, \citenamefont {Style}, \citenamefont
  {Steinmetz} \emph {et~al.}}]{matsuzawa2026metabolites}%
  \BibitemOpen
  \bibfield  {author} {\bibinfo {author} {\bibfnamefont {T.}~\bibnamefont
  {Matsuzawa}}, \bibinfo {author} {\bibfnamefont {K.}~\bibnamefont {Varma}},
  \bibinfo {author} {\bibfnamefont {T.}~\bibnamefont {Bate}}, \bibinfo {author}
  {\bibfnamefont {C.}~\bibnamefont {Lorenz}}, \bibinfo {author} {\bibfnamefont
  {K.}~\bibnamefont {Larina}}, \bibinfo {author} {\bibfnamefont
  {J.}~\bibnamefont {Bauermann}}, \bibinfo {author} {\bibfnamefont
  {D.}~\bibnamefont {Matthias}}, \bibinfo {author} {\bibfnamefont
  {T.}~\bibnamefont {Grubic}}, \bibinfo {author} {\bibfnamefont {R.~W.}\
  \bibnamefont {Style}}, \bibinfo {author} {\bibfnamefont {M.~O.}\ \bibnamefont
  {Steinmetz}}, \emph {et~al.},\ }\bibfield  {title} {\bibinfo {title}
  {Metabolites shift equilibria of biomolecular condensates},\ }\href@noop {}
  {\bibfield  {journal} {\bibinfo  {journal} {bioRxiv}\ ,\ \bibinfo {pages}
  {2026}} (\bibinfo {year} {2026})}\BibitemShut {NoStop}%
\bibitem [{\citenamefont {Rogers}(2010)}]{Rogers_2010}%
  \BibitemOpen
  \bibfield  {author} {\bibinfo {author} {\bibfnamefont {T.}~\bibnamefont
  {Rogers}},\ }\bibfield  {title} {\bibinfo {title} {Universal sum and product
  rules for random matrices},\ }\bibfield  {journal} {\bibinfo  {journal}
  {Journal of Mathematical Physics}\ }\textbf {\bibinfo {volume} {51}},\ \href
  {https://doi.org/10.1063/1.3481569} {10.1063/1.3481569} (\bibinfo {year}
  {2010})\BibitemShut {NoStop}%
\bibitem [{\citenamefont {Barab{\'a}s}\ \emph {et~al.}(2017)\citenamefont
  {Barab{\'a}s}, \citenamefont {Michalska-Smith},\ and\ \citenamefont
  {Allesina}}]{barabas2017self}%
  \BibitemOpen
  \bibfield  {author} {\bibinfo {author} {\bibfnamefont {G.}~\bibnamefont
  {Barab{\'a}s}}, \bibinfo {author} {\bibfnamefont {M.~J.}\ \bibnamefont
  {Michalska-Smith}},\ and\ \bibinfo {author} {\bibfnamefont {S.}~\bibnamefont
  {Allesina}},\ }\bibfield  {title} {\bibinfo {title} {Self-regulation and the
  stability of large ecological networks},\ }\href@noop {} {\bibfield
  {journal} {\bibinfo  {journal} {Nature ecology \& evolution}\ }\textbf
  {\bibinfo {volume} {1}},\ \bibinfo {pages} {1870} (\bibinfo {year}
  {2017})}\BibitemShut {NoStop}%
\bibitem [{\citenamefont {Pastur}(1972)}]{pastur1972spectrum}%
  \BibitemOpen
  \bibfield  {author} {\bibinfo {author} {\bibfnamefont {L.~A.}\ \bibnamefont
  {Pastur}},\ }\bibfield  {title} {\bibinfo {title} {On the spectrum of random
  matrices},\ }\href@noop {} {\bibfield  {journal} {\bibinfo  {journal}
  {Teoreticheskaya i Matematicheskaya Fizika}\ }\textbf {\bibinfo {volume}
  {10}},\ \bibinfo {pages} {102} (\bibinfo {year} {1972})}\BibitemShut
  {NoStop}%
\bibitem [{\citenamefont {Cox}\ and\ \citenamefont
  {Matthews}(2002)}]{cox2002exponential}%
  \BibitemOpen
  \bibfield  {author} {\bibinfo {author} {\bibfnamefont {S.~M.}\ \bibnamefont
  {Cox}}\ and\ \bibinfo {author} {\bibfnamefont {P.~C.}\ \bibnamefont
  {Matthews}},\ }\bibfield  {title} {\bibinfo {title} {Exponential time
  differencing for stiff systems},\ }\href@noop {} {\bibfield  {journal}
  {\bibinfo  {journal} {Journal of Computational Physics}\ }\textbf {\bibinfo
  {volume} {176}},\ \bibinfo {pages} {430} (\bibinfo {year}
  {2002})}\BibitemShut {NoStop}%
\end{thebibliography}%

\end{document}